\documentclass[aps,showpacs,floatfix,nofootinbib,prc,12pt]{revtex4}
\usepackage{graphicx}
\usepackage{epsf}
\usepackage{wrapfig}
\usepackage{epsfig}
\usepackage{sublabel}
\usepackage{verbatim}
\usepackage{amsmath}
\usepackage{color}
\usepackage{soul}
\usepackage{ulem}
\usepackage{multirow}

\def\b0{{\mbox{\boldmath$0$}}}
\def\Vec#1{\mbox{\boldmath $#1$}}

\def\beq{\begin{equation}}
\def\eeq{\end{equation}}
\def\nn{\nonumber}
\def\beqy{\begin{eqnarray}}
\def\eeqy{\end{eqnarray}}

\def\mh{\hat{\mathcal{O}}}

\def \b #1{ {\bf #1}}
\newcommand{\be}{\begin{eqnarray}}
\newcommand{\ee}{\end{eqnarray}}

\def \b #1{ {\bf #1}}
\def \b #1{ {\bf #1}}
     \font\tenbifull=cmmib10 scaled 1200 % bold math italic
     \font\tenbimed=cmmib9
     \font\tenbismall=cmmib7
\mathchardef\bbkappa="7114 \mathchardef\bbrho="711A
\mathchardef\bbsigma="711B \mathchardef\bbtau="711C
\mathchardef\bbvarrho="7125 \mathchardef\bbvarsigma="7126
\mathchardef\bbxi="7118

\begin{document} \vskip 2mm
\date{\today}\vskip 2mm
%------------------------------------------------------------------------------%
\title{Universality of many-body two-nucleon momentum distributions: the correlated nucleon  spectral function of complex nuclei  revisited}
\author{Claudio Ciofi degli Atti$^{1}$}
\email{ciofi@pg.infn.it}
%\author{Chiara Benedetta Mezzetti$^{2}$}
%\email{<chiara.mezzetti@unipg.it>}
\author{\mbox{Hiko Morita$^2$}}
\email{hiko@webmail.sgu.ac.jp} \affiliation{ $^1$Istituto
Nazionale di Fisica Nucleare,
 Sezione di Perugia,\\
 c/o Department of Physics and Geology, University of Perugia,
  Via A. Pascoli, I-06123, Perugia, Italy\\
%$^2$NiPS Laboratory, Department of Physics and Geology, University
%of Perugia,
 % Via A. Pascoli, I-06123, Perugia, Italy\\
$^2$\mbox{Sapporo Gakuin University, Bunkyo-dai 11, Ebetsu
069-8555, Hokkaido, Japan}} \vskip 2mm
\begin{abstract}

\noindent{\bf Background}:
    The nuclear spectral function  is a fundamental  quantity
    which describes the mean-field and short-range correlation dynamics of nucleons
    embedded in the nuclear medium; its knowledge is a prerequisite for the interpretation
    of various electro-weak  scattering processes off nuclear targets aimed at providing  fundamental information on strong and weak interactions.
     Whereas in the  case
    of the three-nucleon  and, partly, the four-nucleon systems,
    the spectral function can be calculated  {\it ab-initio}
     within a non-relativistic many-body
    Schroedinger approach,
    in the case of complex nuclei only  models of the correlated, high momentum part
    of the spectral function are available so far.

\noindent {\bf Purpose}: The purpose of this paper is to present a new approach such that
 the spectral function   for a specific nucleus can be obtained
  from a reliable many-body calculation based upon realistic NN interactions, thus
  avoiding approximations leading to  adjustable
parameters.

\noindent{\bf Methods}: The expectation value of the nuclear many-body Hamiltonian, containing realistic nucleon-nucleon interaction of the
Argonne  family, is evaluated variationally by a normalization conserving linked-cluster expansion  and the resulting many-body correlated
wave functions are used to calculate the one-nucleon  and  the two-nucleon momentum distributions;
by analyzing the high momentum behavior of
the latter, the spectral function can be expressed in terms of a transparent convolution formula
involving the relative and center-of-mass
(c.m.) momentum distributions in specific regions of removal energy E and momentum k.

\noindent{\bf Results}: It is found  that as a consequence of the
 factorization of the many-body
wave functions
at short inter-nucleon separations,
 the high momentum behavior of the
 two-nucleon momentum
distributions in $A=3, 4, 12, 16, 40$ nuclei factorizes, at proper values of the
relative and c.m. momenta, into the c.m. and relative momentum distributions,
with the latter exhibiting a universal A-independent
character.  By exploiting the factorization property,  it is found that
the correlated part of the spectral function  can be expressed in terms
of a convolution formula depending  upon the many-body relative and c.m.  momentum distributions of a nucleon pair.

\noindent {\bf Conclusions}:  The obtained convolution spectral function of
 the three-nucleon systems, featuring both two-and three-nucleon short-range correlations,
perfectly agrees  in a wide range of momentum and removal energy
with the {\it ab-initio} spectral function, whereas in the case of complex nuclei
the integral of the obtained spectral functions ({\it the momentum sum rule}) reproduces with high accuracy
the high momentum part of the one-nucleon momentum distribution,
 obtained independently from the Fourier transform of the non-diagonal
one-body density matrix. Thus, the convolution spectral function we have obtained appears to indeed be a
realistic microscopic, parameter-free quantity
governed by the features of the underlying two-nucleon interactions.
\end{abstract}
 \maketitle
\newpage
%
% SLIDE 1
%
\section{INTRODUCTION: THE NUCLEON SPECTRAL FUNCTION} \label{Sec:1}
The hole spectral function (SF) of nucleon $N_1$,  $P_A^{N_1}({\bf k}_1,E)$ is an
important quantity playing a
 relevant role
in the interpretation of various types of scattering processes off nuclei,
in particular the electro-weak ones; as it
is well known,
it represents the
 joint probability that when  nucleon \lq \lq $N_{1}$\rq \rq
 (usually called the {\it active nucleon})  with momentum ${\bf k}_1$ is removed
 instantaneously
 from the ground state of the nucleus A, the nucleus $(A-1)$ (usually called
  the {\it spectator nucleus}) is left in
 the excited state
 $E_{A-1}^*=E-E_{min}$,  where $E$ is the nucleon  removal energy  and
 $E_{min}= M_{A-1}+m_N-M_A=|E_A|-|E_{A-1}|$,  with $E_A$ and $E_{A-1}$   being
  the (negative)
ground-state energy of nuclei $A$ and $A-1$, respectively. The SF,  which takes into account the fact that
 nucleons in
nuclei have not only a momentum distribution, but also a distribution in energy,
is trivially related to a well defined many-body
quantity, namely the two-points Green's function (see e.g. \cite{1:Dickoff}).
 In what follows
 the well known representation
of the  SF will be used, namely
%%%%%%%%%%%%%%%%%%%%%%%%%%%%%%%%%%%%%%%%%%%%%%%%%%%%%%%%%%%%%%%%%%%% EQUATION
\begin{eqnarray}
   P_A^{N_1}( {\bf k}_1 , E ) & = &
      \frac{1}{2 J + 1}
      \sum_{M , \sigma_1}
      \langle
         \Psi_A^{JM}| a_{{\bf k}_1\sigma_1}^{\dag}
         \delta \left ( E - ( {\hat H_A} - E_{A}) \right )  a_{{\bf k}_1\sigma_1}|\Psi_A^{JM}
      \rangle \\
   & = &
      \frac{1}{2 J + 1}
      \sum_{M , \sigma_1}
      \sum\hspace{-0.5cm}\int_f
      \left|
         \langle
            \Psi_{A - 1}^f | a_{{\bf k}_1\sigma_1} | \Psi_{JM}^{A}
         \rangle
      \right|^{2}
      \delta \left ( E - ( E_{A - 1}^{f} - E_{A} ) \right )  \\
   & = &
      \frac{1}{2 J + 1} ( 2 \pi )^{- 3}
      \sum_{M, \sigma_1}\,
      \sum \hspace{-0.5cm}\int_f
      \left|
         \int \mathrm{d} {\bf r}_1
            e^{\mathrm{i} \Vec{k}_1 \cdot \Vec{r}_1}
           \,G_{f}^{M \sigma_1} ( {\bf r}_1 )
      \right|^{2}
      \delta \left ( E - ( E_{A - 1}^{f} - E_{A} ) \right ) ,
   \label{eq6-1}
\end{eqnarray}
where  $a_{{\bf k}_1\sigma_1}^{\dag}$ ($a_{{\bf k}_1\sigma_1}$) is
the creation (annihilation) operator of a nucleon with momentum
${\bf k}_1$ and spin $\sigma_1$, ${\hat H}_A$ is the intrinsic
Hamiltonian of A interacting nucleons, and
 the quantity
 % $G_{f}^{M_J\sigma} ( \Vec{r}_1 )$
%the overlap integral
%
\begin{equation}
   G_{f}^{M \sigma_1} ( \Vec{r}_1 ) =
      \langle
         \chi _{\sigma_1}^{1/2} ,
         \Psi_{A - 1}^{f} (\{ \Vec{x}\}_{A-1} ) |
         \Psi_{A}^{JM} ( {\bf r}_1, \{ \Vec{x}\}_{A-1})\rangle,
   \label{eq6-2}
\end{equation}
which  has been obtained by using the completeness relation for the eigenstates
 of the nucleus $(A-1)$,
 ($\sum_f |\Psi_{A-1}^f\rangle \langle\Psi_{A-1}^f| =
1$), is the overlap integral between the ground state wave function of nucleus $A$,
 $\Psi_A^{JM}$, and the wave functions of
the discrete and all possible continuum eigenfunctions,
  $\Psi_{A - 1}^{f}$ (with eigenvalue $E_{A - 1}^{f}$ = $E_{A - 1}$ + $E_{A -
1}^{f *}$),  of  nucleus  $(A - 1)$;  eventually, $\{{\bf x}\}$ denotes the
set of spin-isospin and radial coordinates.
The angle integrated SF  is normalized according to %(${\bf k}_1\equiv{\bf k}, |{\bf k}| \equiv k$)
\begin{equation}
 4\,\pi\, \int P_A^{N_1}( k_1, E )\,k_1^2\,d\,{ k}_1 \mathrm{d}E = Z(N),
   \label{eq6-3}
\end{equation}
where N(Z) denotes the number of proton (neutron) in the nucleus. The integral over the removal energy of the SF (the {\it momentum sum rule}) provides the one-nucleon momentum
 distribution
\begin{equation}
n_A^{N_1}({\bf k}_1) = \int P_A^{N_1}( {\bf k}_1, E )\,d\,E,
   \label{eq6-3a}
\end{equation}
 which is linked  to the two-nucleon momentum distribution
 $n_A^{N_1N_2}({\bf k}_1,{\bf k}_2)$, a quantity to be used in what follows,  by
the relation ($N_1\neq N_2$)
%%%%%%%%%%%%%%%%%%%%%%%%%% E Q
\beqy
\hspace{-1cm}n_A^{N_1}({\bf k}_1)=\frac{1}{A-1}
 \left[\int n_A^{N_1N_2}({\bf k}_1,{\bf k}_2)\,d\,{\bf k}_{2}
+2\int n_A^{N_1N_1}({\bf k}_1,{\bf k}_2)\,d\,{\bf k}_{2}\right].
\label{enne1_general}
\eeqy
%%%%%%%%%%%%%%%%%%%%%%%%%% E Q
The one- and two-nucleon momentum distributions are defined  as
follows
%%%%%%%%%%%%%%%%%%%%%%%%%%%%%%%%%%%%%%%%%%%%%%%%%%%%%%%%%%%%%%%%%%%%%%%%%%%%%%%%%BEGIN EQUATION
\beqy
n_A^{N_1}(\Vec{k}_{1})=\,\frac{1}{(2\pi)^3}\int
d\Vec{r_1}\,d\Vec{r_1}^\prime\,
e^{i\,\Vec{k}_{1}\cdot\left(\Vec{r}_1-\Vec{r}_1^\prime\right)}
\,\rho_A^{N_1}(\Vec{r}_1
;\Vec{r}_1^\prime),
\label{1NMomentum}
 \eeqy
 and
\beqy
n^{N_1N_2}_{A}(\Vec{k}_{1},\Vec{k}_{2})=
\,\frac{1}{(2\pi)^6}\int
d\Vec{r_1}\,d\Vec{r_2}\,d\Vec{r_1}^\prime\,d\Vec{r_2}^\prime\,
e^{i\,\Vec{k}_{1}\cdot\left(\Vec{r}_1-\Vec{r}_1^\prime\right)}\,
e^{i\,\Vec{k}_{2}\cdot\left(\Vec{r}_2-\Vec{r}_2^\prime\right)}\,
\rho^{N_1N_2}_{A}(\Vec{r}_1,\Vec{r}_2;\Vec{r}_1^\prime,\Vec{r}_2^\prime) ,
\label{2Nmomdis}
\eeqy
%%%%%%%%%%%%%%%%%%%%%%%%%%%%%%%%%%%%%%%%%%%%%%%%%%%%%%%%%%%%%%%%%%%%%%%%%%%%%%%%%END EQUATION
with the one- and two-nucleon  non-diagonal density matrices,  $\rho_A^{N_1} (\Vec{r}_1
;\Vec{r}_1^\prime)$ and $\rho^{N_1N_2}_{A}(\Vec{r}_1,\Vec{r}_2;\Vec{r}_1^\prime,\Vec{r}_2^\prime)$,
being
%%%%%%%%%%%%%%%%%%%%%%%%%%%%%%%%%%%%%%%%%%%%%%%%%%%%%%%%%%%%%%%%%%%%%%%%%%%%%%%%%%% E Q U A T I O N
\beqy
\rho_{A}^{N_1}(\Vec{r}_1;\Vec{r}_1^\prime)=
\int \psi_A^{JM \,*}(\Vec r_{1},\Vec r_{2},\Vec r_{3}...,\Vec{r}_A)\,
\hat{P}_{N_{1}}(1)
\psi_A^{JM}(\Vec r_{1}^{\prime},\Vec r_{2},\Vec
r_{3},...,\Vec{r}_A)\,\delta\Big(\sum^A_{i=1}\Vec{r}_i\Big)\,
\prod\displaylimits_{i=2}^A d\Vec{r}_i\,,
\eeqy
\beqy
\hskip -0.5cm
\rho_{A}^{N_1N_2}(\Vec{r}_1,\Vec{r}_2;\Vec{r}_1^\prime,\Vec{r}_2^\prime)&=&
\int \psi_A^{JM \,*}(\Vec r_{1},\Vec r_{2},\Vec r_{3}...,\Vec{r}_A)\,
\hat{P}_{N_{1}}(1) \hat{P}_{N_{2}}(2)
\psi_A^{JM}(\Vec r_{1}^{\prime},\Vec r_{2}^{\prime},\Vec
r_{3},...,\Vec{r}_A)\, \nonumber\\
&\times& \delta\Big(\sum^A_{i=1}\Vec{r}_i\Big)\,
\prod\displaylimits_{i=3}^A d\Vec{r}_i\, ,
\eeqy
where $\hat{P}_{N}(i)$ is a projection operator on particle $N$.
Unless differently stated, the following normalizations
 will be used in the rest of the paper
\beqy &&\int n_A^{N_1}(\Vec{k}_{1})  d \Vec{k}_1= Z {\Big
|}_{N_1=p}= N {\Big |}_{N_1=n} , \label{Norma_onebody} \eeqy
\beqy \int n_A^{N_1N_2}(\Vec{k}_{1},\Vec{k}_{2})  d \Vec{k}_1\, d \Vec{k}_2
=
\frac{Z(Z-1)}{2}\,{\Big |}_{N_1=N_2=p}\, =\frac{N(N-1)}{2}{\Big |}_{N_1=N_2=n}\,\, =ZN\,{\Big |}_{N_1=p,N_2=n} ,\label{Norma_partial}
\eeqy
with
\beqy \sum_{N_1N_2}\int n_A^{N_1N_2}(\Vec{k}_{1},\Vec{k}_{2})  d \Vec{k}_1\, d \Vec{k}_2 =\sum_{N_1N_2}\int
\rho_{A}^{N_1N_2}(\Vec{r}_1,\Vec{r}_2) d \Vec{r}_1\, d \Vec{r}_2= \frac{A(A-1)}{2}\,. \label{Norma_total} \eeqy
It can be seen that the
SF and  the one- and two-nucleon  momentum distributions
have to satisfy  simultaneously
 Eq. (\ref{eq6-3a}),  and   Eq. (\ref{enne1_general}). However,
 whereas the calculation of the momentum distributions requires
only the knowledge of the ground-state wave functions, the calculation of the  SF requires
 the knowledge of
both the ground-state wave function of nucleus $A$ and
the entire spectrum of wave functions of the nucleus $A-1$. It is for this  reason that
the SF has  been calculated exactly ({\it ab-initio})
only in the case of the three-nucleon systems  (see \cite{PkE_HELIUM} and \cite{Ciofi_Kaptari_Spectral}),
and partly four-nucleon system \cite{Hiko_Suzo},
whereas   in the case of complex nuclei only models
 can be produced. It should be stressed here, that one of the
 basic requirement for the validity of these models  of the SF
 is the following: when they are integrated in the momentum sum rule (Eq. (\ref{eq6-3a})),
they have to provide the momentum distribution calculated independently by Eq.(\ref{1NMomentum}).
 If short-range correlations (SRC) are taken into account the angle-integrated nucleon SF
 is usually represented in the
 following form \cite{Spec_SM} ($|{\bf k}_1| \equiv k_1 \equiv k$)\footnote{Different
 but equivalent notations are used by different Authors e.g. $P^{N_1}(k,E)=P_{0}^{N_1}(k,E)+P_{1}^{N_1}(k,E)$,
 $P^{N_1}(k,E)=P_{gr}^{N_1}(k,E)+P_{ex}^{N_1}(k,E)$, and others.}
%%%%%%%%%%%%%%%%%%%%%%%%%%%
\beqy
 P_A^{N_1}(k,E)=P_{MF}^{N_1}(k,E)+P_{SRC}^{N_1}(k,E),
\label{DEF_PkE}
 \eeqy
 %%%%%%%%%%%%%%%%%%%%%%%%%%%
  with $P_{MF}^{N_1}$,   describing the mean
field (MF) structure of the nucleus, given by
%%%%%%%%%%%%%%%%%%%%%%%%%%%%
\beqy
P_{MF}^{N_1}(k,E)=\frac{1}{4 \pi} \sum_{\alpha < \alpha_F}
A_{\alpha}n_{\alpha}(k)\,\delta(E-|\epsilon_{\alpha}|),
\label{Spec_SM}
\eeqy
%%%%%%%%%%%%%%%%%%%%%%%%%%%%
where $A_{\alpha}$ denotes the number of
particles in a pure low-momentum ($k \leq 1-1.5 \, fm^{-1}$)
 shell-model state  below the
Fermi sea,  characterized by  a momentum distribution
 $n_{\alpha}(k)$
 and  spectroscopic factor
 %%%%%%%%%%%%%%%%%%%%%%%%%%%
\beqy
N_{\alpha}=\int_0^{\infty}  n_{\alpha}^{SM}(k)\,k^2 \,d\,k <1.
\label{occ probab}
\eeqy
%%%%%%%%%%%%%%%%%%%%%%%%%%%%
In momentum configuration, the  first term in Eq. (\ref{DEF_PkE}) describes  the low
momentum, partially occupied, ground-state shell-model  components
  below the Fermi level,
 whereas the second term  describes high momentum components
  created by  SRC,  whose main effect is to deplete the states below the
  Fermi level, creating occupied states above it.
  As already pointed out,  the correlated part
  of the SF cannot be calculated exactly for $A>4$; as a result,  for complex nuclei
  essentially two models  of the correlated SF have been developed so far. Both of them
   have the general structure of Eq. (\ref{DEF_PkE}) and treat  the uncorrelated part in the same way,
   but different models  are used for
    the correlated part $P_{SRC}^{N_1}(k,E)$:
 in the first model  \cite{Omar_PkE}
the calculated high momentum components in  nuclear matter \cite{Nucl_Matt}
 are  used for  finite nuclei via
 the local
density approximation (LDA), whereas  in the second model
\cite{CiofidegliAtti:1995qe} the
high momentum components  in the nuclear ground-state  arise
from a universal property of the ground-state wave function,
namely its factorization into  short-range and  long-range parts in configuration space,
arising
whenever  a pair of nucleons  is located in the
region of NN interaction dominated by
SRC; in this case  the SF
is expressed in terms of quantities peculiar  for the
given nucleus, namely the center-of-mass (c.m.) and relative momentum distributions of a correlated nucleon pair.
The first model has been intensively and successfully used
in the description of electro-weak processes, in particular in
neutrino scattering off nuclei (see e.g. \cite{Omar_Neutrino}),
 whereas the second one was employed
(see e.g. Ref.\cite{Phys_Rep})
in the analysis  of recent experimental data  on SRC \cite{SRC_exp},
in the interpretation
of deep inelastic scattering \cite{EMC} and
in the extraction of the  nucleon structure
 functions from DIS off nuclei \cite{Kulagin_Petti}. The aim of the present paper is to illustrate
 a novel approach which extends the model of Ref. \cite{CiofidegliAtti:1995qe}   leading to an improved
 realistic microscopic convolution model of   the SF of complex nuclei.

%%%%%%%%%%%%%%%%%%%%%%%%%%%%%%%%%%%%%%%%%%%%%%%%%%%%%%%%%%%%%%%%%%%%%%%%%%%%%%%%%%%%%%%%%%%%%%%%%%%S E C T I O N 4
 \section{Factorization of the many-body nuclear wave functions at
 short relative distances   and
 the correlated momentum distributions}\label{Sec:2}
%%%%%%%%%%%%%%%%%%%%%%%%%%%%%%%%%%%%%%%%%%%%%%%%%%%%%%%%%%%%%%%%%%%%%%%%%%%%%%%%%%%%%%%%%%%%%%%%%%%%%%%%%
\subsection{Factorization: the fundamental property of
 the nuclear wave function at short inter-nucleon ranges}
The assumption  of wave function factorization at short inter-nucleon ranges
is a concept that has been frequently used in the past as a physically sound
 approximation of the unknown nuclear wave function,  mainly to explain
certain classes of medium-energy  experiments (see e.g.
\cite{Fact_Old}),
without providing however any evidence
 of its  quantitative  validity,   due to the lack, at that time, of realistic solutions of the nuclear
  many-body problem. These,
 however, became recently available and  the
 validity of the factorization property could be  checked. As a matter of fact, in the case of
 {\it ab initio} wave functions of few-nucleon systems \cite{CiofidegliAtti:2010xv} the factorization property of
  the wave functions has been demonstrated to hold, and the same was shown to occur in the case of nuclear matter
\cite{Baldo:1900zz}, treated within the Brueckner-Bethe-Goldstone (BBG)
theory \cite{BBG}; moreover,  the general validity  of the factorization property has  also been demonstrated
 in several recent papers
\cite{Weiss:2015mba}. The first approach to employ factorization in order to obtain the
SF  appeared in Ref. \cite{CiofidegliAtti:1995qe};  there indeed
 it has been assumed
   that at short inter-nucleon
 relative distances  ${\bf r}_{ij} = {\bf r}_{i}-{\bf r}_{i}$, much shorter than the center-of-mass coordinate
 ${\bf R}_{ij} = [{\bf r}_{i}+{\bf r}_{i}]/2$ the nuclear wave function
 \beqy
  \hskip-0.5cm  \Psi_A^{JM}(\{\Vec r\}_A) = {\mathcal{\hat A}}\Big\{
\sum_{n,m,f_{A-2}}a_{m,n,f_{A-2}}\Big[ \Big[ \Phi_n(\Vec x_{ij},\Vec r_{ij})\oplus\chi_m(\Vec R_{ij})\Big]
\oplus\Psi_{f_{A-2}}(\{\Vec x \}_{A-2},\{\Vec r \}_{A-2})\Big] \Big\},
\label{wf_exact}
\eeqy
 can be written as follows (see also Ref. \cite{Weiss:2015mba}) \footnote{In Ref \cite{CiofidegliAtti:1995qe} it has been assumed
 that the
 c.m. of the pair moves in $0s$ state implying that factorization occurs only
 when the c.m momentum is very small ($K_{c.m.} \leq 1\, fm^{-1}$)
  with the high momenta being due only to the correlated pairs; as we shall see in what follows factorization can occur also when
 $K_{c.m.}$ is not necessary very low, provided $|{\bf k}_{rel}| >> |{\bf K}_{c.m.}|$.}
\beqy
\hskip-0.7cm
\lim_{r_{ij}<<R_{ij}}\Psi_A^{JM}(\{\Vec r\}_A)\simeq\mathcal{\hat A}\Big\{\chi_{c.m.}(\Vec R_{ij})\sum_{n,f_{A-2}}a_{n,f_{A-2}}\Big[ \Phi_n(\Vec x_{ij},\Vec r_{ij})
 \oplus\Psi_{f_{A-2}}(\{\Vec x\}_{A-2},\{\Vec r \}_{A-2})\Big] \Big\}.
\label{wf_factor}
 \eeqy
%%

 %\cite{CiofidegliAtti:1995qe,Weiss:2015mba}.
 %%
 In Eqs.
(\ref{wf_exact}) and (\ref{wf_factor}):
 {i) $\{\Vec r\}_A$ and $\{\Vec r\}_{A-2}$ denote the set of radial coordinates of nuclei $A$ and $A-2$,
respectively; (ii)   $\Vec r_{ij}$ and $\Vec R_{ij}$
 are the relative and c.m. coordinate of the nucleon pair $ij$, described, respectively, by a short-range
relative  wave function $\Phi_n$ and the c.m. wave function
$\chi_{c.m.}$; iii) $\{\Vec x \}_{A-2}$ and  $\Vec
x_{ij}$ denote the set of  spin-isospin coordinates
of the nucleus  $(A-2)$ and  the pair $(ij)$.
Placing Eq. (\ref{wf_factor}) in the definition of the
two-nucleon momentum
distribution (Eq.(\ref{2Nmomdis})) and changing variables from ${\bf k}_1$, ${\bf k}_2$
to ${\bf k}_{rel}=({\bf k}_1- {\bf k}_2)/2$ and  ${\bf K}_{c.m.}={\bf k}_1+ {\bf k}_2$,
   the following expression of the
two-nucleon momentum distributions is obtained in the region of factorization \cite{2:Alvioli:2016}
%%%%%%%%%%%%%%%%%%%%%%%%%% E Q
\beqy
n^{N_1N_2}_{A}(\Vec{k}_{1},\Vec{k}_{2})&=&n^{N_1N_2}_{A}({k}_{1},{k}_{2},\theta_{12})
=n^{N_1N_2}_{A}({k}_{rel},{K}_{c.m.},\theta)\nonumber\\
&\simeq&n_{c.m.}^{N_1N_2}({K}_{cm}) \,n_{rel}^{N_1N_2}(k_{rel}) ,
\label{2Nmomdis_fact}
\eeqy
%%%%%%%%%%%%%%%%%%%%%%%%%%% E Q
 which is the basic results underlying the short-range structure of nuclei, namely {\it at high values
 of ${\bf k}_{rel}
 >> {\bf K}_{c.m.}$ (${\bf r}_{rel} << {\bf R}_{c.m.}$) the  momentum
 distribution of two correlated nucleons
factorizes into the relative and c.m. momentum distributions, i.e.  no longer depends upon
 angle $\theta$ between ${\bf k}_{rel}$ and ${\bf K}_{c.m.}$}. In other words, when
  SRC are at work,
 the relative and c.m.  motions
 are decoupled.
 A systematic analysis
 of factorization  for nuclei with $A= 3, 4, 12, 16, 40$
 has been presented in Ref. \cite{2:Alvioli:2016} and  the results of this paper
 allowed one to pick up the region of variation
 of the relative and c.m.  momentum distributions where factorization takes place.
 This is a relevant achievement,   for it allows us to
 obtain the SF  in this region free of any adjustable parameter.
  Indeed
 the exact relation between one- and two-nucleon momentum distributions given by Eq. (\ref{enne1_general})
can be expressed,
 \textit{in the factorization
region}, in terms
 of the following convolution formula
 (${\bf k}_{rel} =[{\bf k}_1 -{\bf k}_2]/2={\bf k}_1-{\bf K}_{c.m.}/2$)

%%%%%%%%%%%%%%%%%%%%%%%%% E Q
\beqy
n_A^{N_1}({\bf k}_1)&\simeq& \left[\int
n_{rel}^{N_1N_2}(|{\bf k}_1-\frac{{\bf K}_{c.m.}}{2}|)
n_{c.m.}^{N_1N_2}({\bf
K}_{c.m.})\, d\,{\bf K}_{c.m.}\right.\nonumber\\
 &+& \left.2\int n_{rel}^{N_1N_1}(|{\bf k}_1-\frac{{\bf K}_{c.m.}}{2}|)
  n_{c.m.}^{N_1N_1}({\bf K}_{c.m.})\, d\,{\bf K}_{c.m.}\right]\,d\,{\bf K}_{c.m.} \equiv n_{SRC}^{N_1}({\bf k}_1).
\label{ennep}
\eeqy
%%%%%%%%%%%%%%%%%%%%%%%%%%% E Q
This represents the correlated momentum distributions which will be used in Section \ref{Sec:4} to obtain the correlated SF.
   Before that we will discuss in the next Section the situation concerning the feasibility
    of reliable  many-body calculations based upon realistic models
    of the NN interaction, providing
   parameter-free  ground-state
   wave functions which are necessary to produce the c.m. and relative momentum distributions.

%%%%%%%%%%%%%%%%%%%%%%%%%%%%%%%%%%%%%%%%%%%%%%%%%%%%%%%%%%%%%%%%%%%%%%%%%%%%%%%%%%%%% S E C T I O N
\section{Many-body calculations of the one-nucleon
and two-nucleon momentum distributions} \label{Sec:3}
%%%%%%%%%%%%%%%%%%%%%%%%%%%%%%%%%%%%%%%%%%%%%%%%%%%%%%%%%%%%%%%%%%%%%%%%%%%%%%%%%%%%%%%%%%%%%%%%%%%

\subsection{The realistic many-body approach to the ground-state of nuclei}
During the last few years the calculation of the ground-state property of few-nucleon systems and light nuclei (binding energy and radii, charge density and momentum distributions)
has  reached
a high degree of sophistication  so that  quantities like
 Eqs. (\ref{1NMomentum}) and (\ref{2Nmomdis}) can be
 calculated with ground-state wave functions
 $\Psi_A^{JM}(\{{\bf x}\})$ which are realistic solutions
 of the non-relativistic Schroedinger equation
\be
\hspace{-0.3cm} \left[
      \sum_i \,\frac{\hat{\bf p}_i^2}{2\,m_N}\,+\,\sum_{i<j}
      \,\hat{v}_2({\bf x}_i,{\bf x}_j)+\sum_{i<j<k}
      \hat{v}_3({\bf x}_i,{\bf x}_j,{\bf x}_k) \right]
      \,\Psi_A^f(\{{\bf x}\}_A)=
E_A^f\,\Psi_A^f(\{{\bf x}\}_A).
      \label{Schroedinger}
      \ee
 Here   $\{{\bf x}\}_A \equiv
\{{\bf x}_1,{\bf x}_2,{\bf x}_3,\,\dots,\, {\bf x}_A$\} denotes
the set of A generalized coordinates (the spatial coordinates
satisfying  the condition $\sum_{i=1}^A \Vec{r}_i =0$),
 $f$ stands for
 the complete set of quantum numbers of state $f$ and,
eventually, ${\hat v}_2$ and ${\hat v}_3$  are realistic models of two-nucleon (2N)
 and
three-nucleon (3N) interactions. In what follows we will be mainly
interested in the ground-state wave function
 $\Psi_A^{f=0}\equiv
\Psi_0$. Once the interactions are fixed,
Eq. (\ref{Schroedinger}) should be solved {\it ab initio}, i.e.
exactly, which is possible only the case of few-nucleon systems with $A=3,4$;  for $A>4$ {\it ab initio}
solutions cannot yet be found,  and only approximate
solutions, mostly based on the  variational principle, are available.  Eq.
(\ref{Schroedinger}) has been solved within
various  many-body
approaches
 using    2N interactions which  explain two-nucleon
  bound and scattering
 data and, considering, also 3N interactions, which are
 introduced to explain the properties of the 3N bound states.
 In these calculations advanced forms of the NN interaction are  provided by the so called Argonne
 family, in which case they have the following general form  \cite{Wiringa:1994wb}

\beqy {\it v}({\bf x}_i,{\bf x}_j)=\sum_{n=1}^{n_{max}} {\it
v}^{(n)}(r_{ij}) {\mathcal O}_{ij}^{(n)}, \label{potential} \eeqy
where ${\bf x}_k\equiv\{{\bf r}_k,{\bf s}_k,{\bf t}_k \}$
  denotes the set of  nucleon radial, spin and isospin
coordinates, ${\mathcal O}_{ij}^{(n)}$
is a proper operator
depending upon the orbital, spin and isospin momenta, and
$n_{max}=18$; in the case of purely central interaction
one has
${\mathcal O}_{ij}^{(n=1)}=1$ and
${\mathcal O}_{ij}^{(n>1)}=0$, whereas in the realistic case the most
important operators are as follows
\beqy
 \mh^{(1)}_{ij}&\equiv&\mh^{c}_{ij}\,=\,1\,\hspace{6.2cm}
\mh^{(2)}_{ij}\,\equiv\,\mh^{\sigma}_{ij}\,=\,\Vec{\sigma}_i\cdot\Vec{\sigma}_j\,\nn\\
\mh^{(3)}_{ij}&\equiv&\mh^{\tau}_{ij}\,=
\,\Vec{\tau}_i\cdot\Vec{\tau}_j\,\hspace{5.3cm}
\mh^{(4)}_{ij}\,\equiv\,\mh^{\sigma\,\tau}_{ij}\,=
\,(\Vec{\sigma}_i\cdot
\Vec{\sigma}_j)\,(\Vec{\tau}_i\cdot\Vec{\tau}_j)\,\nn\\
 \mh^{(5)}_{ij}&\equiv&\mh^{t}_{ij}\,=\,\hat{S}_{ij}\,
 \hspace{5.9cm}
\mh^{(6)}_{ij}\,\equiv\,\mh^{t\,\tau}_{ij}\,=
\,\hat{S}_{ij}\, (\Vec{\tau}_i\cdot\Vec{\tau}_j)
\label{n2_rel},
\eeqy
where $\hat{S}_{ij}$ is the tensor operator.
Using such an NN potential, supplemented by 3N forces, {\it
ab-initio} solutions of the 3-body \cite{Kievsky:1992um}
and 4-body \cite{Akaishi:1987} nuclei,
have been obtained. As for $A>4$ nuclei
realistic
  ground-state wave functions  are available from  variational calculations, i.e. from the minimization of the expectation
value of realistic non
relativistic Hamiltonians, namely
%%%%%%%%%%%%%%%%%%%%%%%% E Q
\be
\langle {\hat H} \rangle = \frac{\langle \Psi_0 | {\hat H} | \Psi_0 \rangle}{\langle \Psi_0 |
\Psi_0 \rangle} \equiv E_A^V \geq E_A^0, \label{2:EV}
\ee
%%%%%%%%%%%%%%%%%%%%%%%% E Q
assuming  the following  correlated wave function as the variational one
%%%%%%%%%%%%%%%%%%%%%%%% E Q
\be
\Psi_0(\{ {\bf x}\}_A)= {{\hat F}(\{ {\bf x}\}_A)}\Phi_0(\{ {\bf
x}\}_A),
\label{2:CBF1}
\ee
where $\Phi_0(\{ {\bf x}\}_A)$  is a mean-field wave function and
%%%%%%%%%%%%%%%%%%%%%%%% E Q
 \beqy
  \hat{F}(\{{\bf x} \}_A)=\hat{{S}_A}\prod_{i<j} \left [ \sum_{n=1}^{n_{max}} f^{(n)}(r_{ij})
  \,
\hat{O}^{(n)}_{ij}\right]
\label{2:Corroper}
 \eeqy
 is a symmetrized (by the operator $\hat{{S}_A}$) product of operators
$\hat{\mathcal{O}}^{(n)}_{ij}$
(the same that appear in the two-nucleon interaction (Eq. (\ref{potential}))
and $f^{(n)}$ is a correlation
which reflects the features of the two-nucleon interaction and
 cures its possible singularities, e.g.
 if only central hard core interactions
are considered, the well known Jastrow form is obtained  \cite{Jastrow}
\be
{\hat F}_J(\{ {\bf x}\}_A)=
\prod_{i<j}f_C(r_{ij}),
\label{2:Jastrow}
\ee
where $f_C(r_{ij})=0$ when $r_{ij}\leq r_c$, if the two-nucleon potential
exhibits  a hard core of radius $r_c$.
  For
complex nuclei
 with $A \leq 12$,  Eq. (\ref{2:EV}) has been evaluated exactly within the Variational Monte Carlo
 (VMC) approach
 \cite{Wiringa:2013ala},
 based upon the numerical evaluation of the multidimensional integrals;
  by this way the VMC ground-state energy and
 wave-functions have been obtained and the momentum distributions were
 accordingly calculated. For $A>12$ the increasing dimension
 of the required integrals related to the non central part of the potential, forbids
  till now the exact evaluation
 Eq. (\ref{2:EV}), so that  some approximations are still necessary.
 In the Cluster Variational Monte Carlo (CVMC) \cite{CVMC_Panda}
  the  contributions arising from
 the central part of the interaction are evaluated exactly with Jastrow-like wave
 functions,
whereas the contributions arising  from the non central part of the interaction were considered only for a
 limited number (five) of
 correlated nucleons; the CVMC has been  recently applied to
  the description of
  $^{16}O$ and $^{40}Ca$ nuclei
 \cite{CVMC_ARGO}.
 Thus due to the heavy numerical
computation efforts required by the  increasing number of
 nucleons,  also CVMC
is  still difficult to perform and various alternative
 methods have been so far
 developed, based, in close analogy with the
  theory of quantum fluids \cite{Fluids}, upon the evaluation of the leading
  contributions  of Eq. (\ref{2:EV}); in particular, the following
  approaches should be mentioned:
 (i) the fermion hypernetted chain method (FHNC),
  where a certain class of contribution
  (the nodal diagrams),  are summed to all orders (see:
  \cite{FHNC,AriasdeSaavedra:2007byz})
  and (ii)
  various  cluster expansion approaches \cite{Cluster_expansions,Ripka}
  in which the expectation value of a given operator is
   rearranged in a series,
  whose zero-th order term is the mean field contribution and the n-th
  order term provides the contribution from $n$ correlated nucleons.
  In this connection let us stress, as it is well known
  and also recently
  recalled \cite{CVMC_ARGO}, that the  procedure
  of considering lowest order terms in the numerator and in the denominator of the expectation value
  of a certain operator  and
  then taking their ratio, should not be pursued  due to the
    presence, both in the numerator and the denominator, of unlinked terms
    which produce the divergence of the ratio with
     increasing number of particles.
    In our approach,  we
  have followed the {\it normalization conserving linked cluster expansion}
  (NCLCE)  developed in Ref. \cite{Ripka},
applied in the case of central interactions  in Ref. \cite{Bohigas} and generalized in  Ref. \cite{2:Alvioli:2005cz} to the case of
realistic interactions and applied to the calculations of the properties of $^{16}$O and $^{40}Ca$.
The main feature of  NCLCE can be illustrated in the simple case of
 the calculation of the expectation value of
a generic operator
$\hat{\cal O}$ and a Jastrow-like wave function , i. e. in the
 case of

\be
  <\hat{\cal O}>=
\frac{<\Psi|\hat{\cal O}|\Psi>}{<\Psi|\Psi>}=
\frac{<\psi_{MF}|\prod f(r_{ij})\hat{\cal O}
 \prod f(r_{ij})|\psi_{MF}>}{<\psi_{MF}|\prod f(r_{ij})^2
|\psi_{MF}>}.
\label{expectation}
\ee
By writing
\be
f(r_{ij})^{2}=1+\eta(r_{ij})
\label{eta}
\ee
and expanding the resulting denominator in Eq. (\ref{expectation}),
$[1+x]^{-1}=1-x+x^2-...$,  it can be shown that the unlinked terms in the numerator exactly
cancel out the ones
arising from the  denominator and a convergent series
expansion containing only linked terms is obtained in the following  form

\be
<\hat{\cal O}>=\langle \psi_{MF}|{\hat{\cal O}}|\psi_{MF} \rangle +
\langle \hat{\cal O}\rangle_1 +
\langle \hat{\cal O} \rangle_2+\dots ...+ \langle \hat{\cal O} \rangle_n +....,
\label{2:Clustexp}
\ee
where the subscripts  denote the number of $\eta_{ij}$ appearing
in the given term, $<\psi_{MF}|\hat{\cal O}|\psi_{MF}>$
represents the MF uncorrelated
contribution and the other
terms represent the contribution from all linked and topologically
 distinct Ivon-Mayer diagrams \cite{Ivon_Mayer},
describing  clusters of correlated nucleons\footnote{Note, in order to avoid confusions, that the first term of Eq. (\ref{2:Clustexp}) is a
pure independent-particle contribution, whereas in the definition of the SF (Eq. (\ref{DEF_PkE}))
the mean-field part $P_{MF}^{N_1}(k,E)$
is renormalized by the spectroscopic factor of the single particle orbits.}. For example the first order term is explicitly written as
\cite{2:Alvioli:2005cz}
\begin{eqnarray}
  \label{NCLCE_Jastrow}
  \langle \hat{\cal O}\rangle_1&=&<\psi_{MF}|\sum_{i<j}\left(f(r_{ij})\hat{\cal O}f(r_{ij})-\hat{\cal O}\right)|\psi_{MF}>  \\ \nonumber
  &-&<\psi_{MF}|\hat{\cal O}|\psi_{MF}><\psi_{MF}|\sum_{i<j}\left(f(r_{ij})^2-1\right)|\psi_{MF}> .
\end{eqnarray}
If the correlation function has the form like Eq. (\ref{2:Corroper}), the above expression is extended to the following form
\begin{eqnarray}
  \label{NCLCE_general}
  \langle \hat{\cal O}\rangle_1&=&<\psi_{MF}|\sum_{i<j}\left(\hat{f}(ij)\hat{\cal O}\hat{f}(ij)-\hat{\cal O}\right)|\psi_{MF}>  \\ \nonumber
  &-&<\psi_{MF}|\hat{\cal O}|\psi_{MF}><\psi_{MF}|\sum_{i<j}\left(\hat{f}(ij)\hat{f}(ij)-1\right)|\psi_{MF}>,
\end{eqnarray}
where
\begin{eqnarray}
  \label{f_ij}
   \hat{f}(ij)\equiv \sum_{n=1}^{n_{max}}f^{(n)}({\bf r}_{ij})\hat{O}_{ij}^{(n)}.
\end{eqnarray}
 The merit of this approach is the full cancelation  of  unlinked clusters contribution,
  which is a prerequisite for any  convergent cluster expansion.  The explicit expressions of the
  one- and two-nucleon non diagonal density matrices at the first order, which include up to clusters of four particles are given in Appendix. They are the basic quantities which are necessary to obtain the one-nucleon and two-nucleon momentum distributions.

  Once  the cluster expansion has been chosen  the problem remains  of the choice
  of the variational parameters  which characterize both the wave function and the correlation functions.
  Indeed these have to be chosen as the ones which minimize the expectation value of
  the Hamiltonian (Eq. (\ref{2:EV})). As far  as the correlation functions are concerned, it is a common
   practice (see e.g. Ref. \cite{AriasdeSaavedra:2007byz}) to obtain their  shape
by the minimization of the Hamiltonian
at lowest order,  obtaining  by this way  Euler-Lagrange equations which  fix
the shape of the correlation functions $f^{(n)}(r)$, according to  the following conditions
\be
f^{(p=1)}(r)=f_c(r)\rightarrow 1 \qquad  \textrm {at} \qquad r \geq d\\
f^{(p>1)}(r) \rightarrow 0 \qquad \,\,\,\, \textrm {as} \qquad
r\rightarrow \infty,
\label{2:asintfp}
\ee
where $d$, the healing distance, representing the distance
beyond which the two body correlated wave function $\psi(12)$ heals to the uncorrelated one
$\phi(12)$,  becomes the general variational parameter of  the expansion together with
the mean-field parameters.
To sum up, there are at the moment realistic many-body wave functions, solutions
of  Eq. (\ref{Schroedinger}), which
 can be used to calculate realistic momentum distributions
and  model SF,
without recurring to parameterized  wave functions not corresponding to the minimization of the
ground-state energy, or model wave functions  containing adjustable parameters. At the same time,
it turns out, as it will be shown in
what follows,  that the approach described above,  namely a parameter-free
NCLCE  can provide, with much less numerical efforts,
  results for the ground-state properties of light and
medium weight nuclei in reasonable agreement with VMC
\cite{Wiringa:2013ala} and CVMC results \cite{CVMC_ARGO}.
 In the next Subsection,   following Ref.
  \cite{2:Alvioli:2016},
we will compare
the results of our approach with the results of various many-body calculations of the
ground-state energy and the one- and two-nucleon momentum
distributions,  whereas in Section \ref{Sec:4},
 following the procedure of
Ref. \cite{3NC_Helium}, we will
 present the results for
 the SF of  complex nuclei.

\subsection{Comparison of our results with the results of VMC and CVMC many-body approaches}
\subsubsection{Binding energies, two-nucleon correlation functions
 and one-nucleon momentum distribution}
In Table \ref{Table1} and Figs. \ref{Fig1}-\ref{Fig4}
 we compare the results of our NCLCE calculations with
 the results
 of other methods, particularly the
 VMC \cite{Wiringa:2013ala} and CVMC  \cite{CVMC_ARGO} ones obtained with similar NN interactions,
omitting and
including 3N forces. In Table \ref{Table1} the values of the ground-state energy
 and r.m.s radii are compared, whereas
 Fig. \ref{Fig1}  shows the two-body densities associated to the
 six correlation functions corresponding to the operators given in Eq. (\ref{n2_rel}).
 An acceptable similarity of  our results with the most advanced CVMC approach can be seen.
 In Fig. \ref{Fig2}  we compare  the one-nucleon momentum distribution of
$^{16}O$ and $^{40}Ca$ we have obtained in Ref. \cite{2:Alvioli:2005cz} with   recent
CVMC results \cite{CVMC_ARGO} and a remarkable agreement is evident \footnote{In previous
 and present calculations we did not include  the 3N
 interaction in Eq. (\ref{Schroedinger}),
  since we considered that the effects of the known 3N forces,
  conceived in order to provide the  missing binding in  $^3$He, obtained  when
   only 2N forces are considered, should not produce large effects on the high momentum
  content of the momentum distribution, as indeed was demonstrated by recent CVMC
  in $^{16}$O and $^{40}$Ca (see Figs. 11, 12 and 13 of Ref. \cite{CVMC_ARGO})}. In Fig. \ref{Fig3}
we also show the results of several different approaches to the momentum distributions
of $^{16}O$. Since, as usually, the momentum distributions are given on a log plot, in Fig.\ref{Fig4}
 we show the quantity
\be
\Delta n(k)=100\frac{n_{x}(k)-n_{CVMC}(k)}{n_{VCVM}(k)}
\label{deviation}
\ee
measuring the percent deviation of the theoretical
momentum distribution of $^{16}O$ shown in Fig. \ref{Fig3},
with respect to the  CVMC
results of Ref. \cite{CVMC_ARGO}, taken as the reference
momentum distributions. From this plot it  can again be seen  that
 our one-nucleon momentum distributions are sufficiently realistic ones.
\subsubsection{Two-nucleon momentum distributions}
In this subsection  we will compare  the
two-nucleon  momentum distributions
calculated within the VMC approach \cite{Wiringa:2013ala}
with the momentum distributions obtained within our NCLCE
approach  \cite{2:Alvioli:2016}.
The two-nucleon momentum distribution is function of three
 variables, namely the relative momentum
  $|\Vec{k}_{\text{rel}}| \equiv k_{rel}$,
 the c.m. momentum
 $|\Vec{K}_{\text{c.m.}}| \equiv K_{c.m.}$ and the angle $\theta$
 between them,
\beqy &&n_A^{N_1N_2}(\Vec{k}_{\text{rel}},
\Vec{K}_{\text{c.m.}})
=n_A^{N_1N_2}({k}_{\text{rel}},{K}_{\text{c.m.}},\theta)\,=\nonumber\\
&&=\frac{1}{(2\pi)^6}\int
d\Vec{r}\,d\Vec{R}\,d\Vec{r}^\prime\,d\Vec{R}^\prime\,
e^{i\,\Vec{K}_{\text{c.m.}}\cdot\left(\Vec{R}-\Vec{R}^\prime\right)}\,
e^{i\,\Vec{k}_{\text{rel}}\cdot\left(\Vec{r}-\Vec{r}^\prime\right)}\,
\rho^{(2)}_{N_1N_2}(\Vec{r},\Vec{R};\Vec{r}^\prime,\Vec{R}^\prime).
\label{2brelc.m.}
\eeqy
Here we will consider  two different momentum distributions namely: the c.m. momentum distribution
%%%%%%%%%%%%%%%%%%%%%%%%%%%%% E Q
\beqy
{n_A^{N_1N_2}(K_{c.m.})}= \int \, n_A^{N_1N_2}(\Vec{k}_{\text{rel}},\Vec{K}_{\text{c.m.}}) d\,\Vec{k}_{\text{rel}}
\label{n2_CM} \,\equiv n^{N_1N_2}_{c.m.}({K}_{\text{c.m.}})\, ,
\eeqy
 shown in Fig. \ref{Fig5}, and
 the relative momentum distribution
 %%%%%%%%%%%%%%%%%%%%%%%%%%%% E Q
\beqy
 {n_A^{N_1N_2}({k}_{\text{rel}})}= \int
\,n_A^{N_1N_2}(\Vec{k}_{\text{rel}},\Vec{K}_{\text{c.m.}})\,d\,\Vec{K}_{\text{c.m.}}
\label{n2_rel_2}\, ,\eeqy
shown in Fig \ref{Fig6}.
It can be seen that  an overall satisfactory agreement does indeed
 occurs between the VMC and the NCLCE approaches.
The general $\theta$-dependent
two-nucleon momentum distribution (Eq. (\ref{2brelc.m.})) has already been presented in
Ref. \cite{2:Alvioli:2016}. In this paper a new plot of this quantity will be
given in the next Section.

\subsubsection{Summary}
An overall agreement of the results of calculations
performed
with VMC and NCLCE approaches has been found as far as
 the one-nucleon and two-nucleon relative and c.m momentum distributions
 of few-nucleon systems and
medium-weight nuclei are concerned.  Such an agreement
 makes us confident that the
  full momentum distributions calculated at different values
  of $K_{c.m}$, $k_{rel}$ and $\theta$, the quantities  which are necessary
  for the production of the nuclear SF, are
  genuine and realistic many-body quantities free of
  any adjustable parameter.

\section{Wave  function factorization and  the many-body convolution formula
 of the correlated
spectral function} \label{Sec:4}
\subsection{The universal factorized  behavior of the two-nucleon momentum
distribution}
In Section \ref{Sec:2} we have demonstrated that if
the two-nucleon momentum distribution
factorize,  the convolution formula of the SRC momentum distributions
is obtained. By plotting the two-nucleon momentum
distributions {\it vs}
$|{\bf k}_{rel}|$ at different fixed values of the c.m momentum
 $|{\bf K}_{c.m.}|$  and og the angle ${\theta}$ between
 the two momenta,
it has indeed  been shown \cite{2:Alvioli:2016}
that at sufficiently high values of
the relative momentum, such that
$|{\bf k}_{rel}|>> |{\bf K}_{c.m}|$,
the two-nucleon momentum distributions  indeed factorize. In order to more
quantitatively identify the factorization regions, in Fig. \ref{Fig7} we show    a  3D plot of the two-nucleon
 momentum distribution
  pertaining to $^4$He at  $\theta =0^o$ and $\theta =90^o$
 (similar results are available for other nuclei).
The factorization regions, i.e. the region where the result at both angles coincide,  can clearly be seen. A further  important feature of factorization,
which was overlooked in
Ref. \cite{2:Alvioli:2016}, but stressed in Ref. \cite{3NC_Helium}, is also
visible:
 factorization is not only valid  in the  region of low c.m. momenta  but also in the region of high c.m. momenta.
 In this respect  it should be stressed that the minimum value  of the relative
   momentum at which
  factorization starts to occur is a function  of the value
  of the c.m. momentum $K_{c.m.}$, namely
  factorization is valid  when
  %%%%%%%%%%%%%%%%%%%%%%%%%% E Q
  \beqy
  k_{rel} \gtrsim k_{rel}^-(K_{c.m.}),
  \label{kameno}
  \eeqy
  with \cite{3NC_Helium}
  %%%%%%%%%%%%%%%%%%%% E Q
  \beqy
  k_{rel}^-(K_{c.m.}) \simeq a+b\,\phi(K_{c.m.}),
  \label{kameno1}
\eeqy
where $a \simeq 1 \, fm^{-1}$ and the function $\phi(K_{c.m.})$ is such that
$\phi(0) \simeq 0$ \footnote{This condition is somewhat softer than that used for $^3$He
 in Ref. \cite{3NC_Helium}.  Indeed we carefully reanalyzed the factorization-condition (Eq. (\ref{kameno1}))
  and
 found that $k_{rel}^{-}=1.0+0.5K_{c.m.}$ is the most accurate one within the linear-$K_{c.m.}$ dependence.
 Thus in the rest of the paper we use this factorization-condition also in the case of
 the  $^3$He SF.}. Since the value of $k_{rel}^-$ depends upon the value of
$K_{c.m.}$,  Eq. (\ref{kameno1})  generates in Eq. (\ref{ennep}) a constraint on the
region of integration over ${\bf K}_{c.m.}$, in that only those values of ${\bf K}_{c.m.}$
  satisfying Eq. (\ref{kameno1})
 have to be considered. For a fixed value of $k_1$ the relation between $k_1$ and $K_{c.m.}$, given by
%%%
\beqy
 k_{rel} =|{\bf
k}_1-\frac{{\bf K}_{c.m.}}{2}| \geq k_{rel}^{-}(K_{c.m.}),
 \label{restriction}
 \eeqy
%%%
 represents the equation  which establishes a constraint on the
  the region of integration over ${\bf K}_{c.m.}$; this region becomes  narrower
 than the region which is obtained  if  the constraint given by Eq. (\ref{restriction}) is disregarded.
 It is worth stressing  that except for Ref. \cite{3NC_Helium}, Eq. (\ref{restriction}) and the resulting constraint were never been considered in the past.

  The independence of the two-nucleon momentum distribution (Eq. (\ref{2brelc.m.})) upon the angle $\theta$ is  direct proof
  that
factorization does occur for both $pn$ and $pp$ SRC pairs, which means that
%%%%%%%%%%%%%%%%%%%%%%%%%%%%% E Q
\beqy
n_A^{N_1N_2}({\bf k}_1,{\bf k}_2)=n_A^{N_1N_2}({ k}_{rel}, { K}_{c.m.}, \theta) \simeq
  n_{rel}^{N_1N_2}({ k}_{rel})\, n_{c.m.}^{N_1N_2}({ K}_{c.m.}).
  \label{factorization}
\eeqy
Moreover, in the case of $pn$ pairs one finds \cite{2:Alvioli:2016}
 %%%%%%%%%%%%%%%%%%%%%%%%%%%%% E Q
\beq
 n_A^{pn}({ k}_{rel},{ K}_{c.m.})\simeq C_A^{pn}
  n_D({ k}_{rel})\,n_{c.m.}^{pn}({K}_{c.m.});
\label{factpn}
 \eeq
 where $n_{D}$ is the deuteron momentum distribution  and $C_A^{pn}$ is a constant depending upon the
 atomic weight and which, together with the integrals of $ n_{D}({ k}_{rel})$ and  $n_{c.m.}^{pn}({ K}_{c.m.})$
 in the proper SRC region,  counts the number of SRC $pn$ pairs in the given nucleus.
 Since the quantities   $n_A^{pn}({k}_{rel},{K}_{c.m.})$,
  $n_D({ k}_{rel})$ and $\,n_{c.m.}^{pn}({K}_{c.m.})$ are genuine many-body quantities, so is the value of $C_A^{pn}$
  given by
\beqy
C_A^{pn}=\frac{n_A^{pn}({ k}_{rel},{ K}_{c.m.})}{ n_D({ k}_{rel})\,n_{c.m.}^{pn}({K}_{c.m.})}\, .
\label{Ci_Apn}
\eeqy
Factorization,  which has  recently been confirmed also in Ref. \cite{Weiss:2015mba},
 stays now on solid grounds, and so is the relation between the
one-nucleon and two-nucleon momentum distributions
 given by Eq. (\ref{ennep}).
Whereas the $pn$ two-nucleon momentum distribution  in the factorization region can be expressed in
terms of the
deuteron momentum distribution, the $pp$ distribution  cannot be related to a known free $pp$
function; nonetheless  they also show a regularity  which is exhibited for
 $^4$He and $^{40}Ca$
in Figs. \ref{Fig8} and 9. These figures demonstrate that
the $k_{rel}$ dependence
 of the $pp$ distribution at various values of $K_{c.m.}$  is governed in the factorization region by a common
 function of  $k_{rel}$, with the amplitude determined by  the value of $K_{c.m.}$.
 Thus if one defines the
quantity
%%%%%%%%%%%%%%%%%%%%%%%%%%%%%% E Q
\beq {\widetilde
n_{rel}}^{pp}(k_{rel})=
\frac{n_{rel}^{pp}(k_{rel},K_{c.m.}=0)}{n_{c.m.}(K_{c.m.}=0)},
\label{nrel_Kcm0}
\eeq
one finds that in the factorization region the $pp$ momentum distribution assumes the following form
%%%%%%%%%%%%%%%%%%%%%%%%%%%%% E Q
\beq
n_{A}^{pp}(k_{rel},K_{c.m.})\simeq {\widetilde
n_{rel}}^{pp}(k_{rel}){n_{c.m.}(K_{c.m.})}, \label{ennepp}
\eeq
which exhibits, as clearly appears from Figs. \ref{Fig8} and 9,
a very good agreement with the exact
calculation.

  We have now at disposal all microscopic many-body quantities to evaluate the
  one-nucleon  SF, namely  Eqs.
  (\ref{n2_CM}), (\ref{factpn}) and (\ref{ennepp}); having at disposal the SF we can calculate back the momentum distributions that, as previously
  stressed,
  has  to coincide with the momentum distribution calculated directly
  with Eq.(\ref{1NMomentum}).

\subsection{The spectral function of A=3, 4, 12, 16, 40}
On the basis of what has been presented in the previous Sections,
the total one-nucleon SF can be written in the
following form \beqy
P_A^{N_1}(k,E)&=&P_{MF}^{N_1}(k,E)+P_{SRC}^{N_1}(k,E) \equiv
P_{conv}(k,E) \label{Spec_fin} \eeqy
where the mean-field contribution $P_{MF}^{N_1}(k,E)$ is given by Eq. (\ref{Spec_SM})
and
  \beqy P_{SRC} ^{N_1}({\bf k}_1,E)&=&
\int n_{rel}^{N_1N_2}(|{\bf k}_1-\frac{{\bf K}_{c.m.}}{2}|)\,
  n_{c.m.}^{N_1N_2}({\bf K}_{c.m.})d\,{\bf K}_{c.m.}\,\nonumber\\
  &\times&\delta
\left( E-E_{thr}^{N_1}- \frac{A-2}{2m_N(A-1)} \left[{\bf k}_1-\frac{(A-1){\bf K}_{c.m.}}{A-2} \right]^2\,
\right)\nonumber\\
&+\,2&\int
n_{rel}^{N_1N_1}(|{\bf k}_1-\frac{{\bf K}_{c.m.}}{2}|)\,
  n_{c.m.}^{N_1N_1}({\bf K}_{c.m.})d\,{\bf K}_{c.m.}\,\nonumber\\
  &\times&\delta
\left( E-E_{thr}^{N_1}- \frac{A-2}{2m_N(A-1)} \left[{\bf k}_1-\frac{(A-1){\bf K}_{c.m.}}{A-2} \right]^2\, \right)
\label{Spec_SRC_fin}
\eeqy
with $N_1 \neq N_2$. Let us remind that $P_{MF}^{N_1}(k,E)$ arises from  the mean field, namely independent
 particle motion,
whereas  $P_{SRC} ^{N_1}({\bf k}_1,E)$ arises from the factorization of the nuclear wave function as
 in Eq. (\ref{wf_factor}), assumed to hold (see also Ref. \cite{CiofidegliAtti:1995qe,Weiss:2015mba})) when
 ${\bf r}_{ij}<<{\bf R}$ (or ${\bf k}_{rel}>> {\bf K}_{c.m.}$), the assumption that leads, in turns, to the factorization of the two-nucleon  momentum and to  Eq. (\ref{Spec_SRC_fin}).

Eq. (\ref{Spec_SRC_fin}) is the {\it convolution formula}
of the correlated part of the SF. It represents  the SF in the
 so-called plane wave approximation (PWA), which  describes the process in which a correlated nucleon
 removed from a correlated pair, leaves the nucleus  without interacting with
 nucleus $A-1$,
 whose excitation energy  is therefore given  by the sum of the
 threshold energy $E_{thr}= |E_A|-|E_{A-1}|$
  plus the relative
 kinetic energy of  the system: \textit{\lq \lq nucleus $(A-2)$- recoiling nucleon of the initially correlated pair \rq \rq }.
 It has been shown in Ref. \cite{3NC_Helium}, on the example
 of the {\it ab-initio} 3N SF \cite{Ciofi_Kaptari_Spectral},
that in a wide range of high values
 of momentum and
 removal energy typical of SRCs,
 the PWA  SF is  practically indistinguishable from the results of
 the Plane Wave Impulse Approximation
 (PWIA) SF in which the exact continuum two-nucleon wave function of the correlated pair
  is taken into account.

Let us now summarize two  main features  of the correlated SF:
\begin{itemize}
\item the correlated SF (\ref{Spec_SRC_fin}) depends upon two  basic ground-state properties of
nuclei, namely
the c.m. and relative $pn$ and $pp$ momentum distributions, two quantities that have been calculated within
advanced and rigorous many-body theories (VMC, NCLCE) so that Eq. (\ref{Spec_SRC_fin}) is a genuine realistic
many-body quantity free of any adjustable parameter.
\item the only model dependence of (\ref{Spec_SRC_fin}) resides in the argument of the energy-conserving
delta function;
 such an approximation
is justified by  the high values of the removal energies characterizing the SRC SF;
 \item it should be stressed that Eq. (\ref{Spec_SRC_fin}) was essentially firstly obtained in
 Ref. \cite{CiofidegliAtti:1995qe} but applied there with phenomenological effective two-nucleon relative
 and c.m. momentum  distributions. We should
 also point out that recently a model SF has been obtained within a relativistic kinematics
  approach \cite{Misak}, leading
 to the result of Ref. \cite{CiofidegliAtti:1995qe} in the non relativistic limit.
\end{itemize}

In Fig. \ref{Fig10} we show the proton and neutron SF of $^3$He, calculated by Eq.(\ref{Spec_SRC_fin}), compared
with the {\it ab-initio} SF of Ref.
\cite{Ciofi_Kaptari_Spectral};
 the SF of $^4$He, $^{12}$C, $^{16}$O and $^{40}$Ca, are shown in
Fig. \ref{Fig11} where the separate contributions of $pp$ and $pn$ SRC are illustrated;
the comparison with the
convolution model of Ref. \cite{CiofidegliAtti:1995qe} is presented in Fig. \ref{Fig12}.
In all of these figure
$k=3.5\, fm^{-1}$. The $k$ and $E$ dependencies of the SF of Eq. (\ref{Spec_fin}) in the case of $^{12}$C are shown in a $3D$ plot
 in Fig. \ref{Fig13}.
Let us comments the main features of these results. Concerning the three-nucleon system
(see also Ref. \cite{3NC_Helium}),
it is very gratifying
to observe a remarkable agreement  of our convolution formula with the {\it ab initio}
results  in a wide range of
 removal energy,
particularly in light of the absence of any adjustable parameter in Eq. (\ref{Spec_SRC_fin});
as for complex nuclei,
 the small contribution of $pp$ SRC with respect to $pn$ SRC, in agreement with
experimental evidences \cite{SRC_exp}, should be stressed; concerning the
differences between the present approach and the approach of
Ref. \cite{CiofidegliAtti:1995qe}, where the convolution formula for
the SF  has been firstly applied,  the following remarks are in order:
%%%%%
\begin{enumerate}
\item both approaches have the same origin  and structure, which is  the convolution formula resulting from
 wave function factorization,  with
the main difference between the two approaches  being related to the relative and c.m. momentum distributions   used in the
convolution formula; indeed in Ref. \cite{CiofidegliAtti:1995qe}, due to the lack of realistic many-body
calculations for complex nuclei, effective momentum distributions for $pp$ and $pn$ have been used, moreover
at that time  the region of factorization, which ensures  the validity of the convolution formula,
was  unknown;
\item the differences between $pp$ and $pn$ momentum distributions, which
 is a prerequisite for extending the convolution
approach  to non-isoscalar asymmetric nuclei, have not been considered in Ref. \cite{CiofidegliAtti:1995qe}, for
the reasons given above;
\item in  Ref. \cite{CiofidegliAtti:1995qe}  only the
soft part  of the c.m. momentum distribution  has been considered and  the constraint on the values of $K_{c.m.}$  was
 disregarded.
\end{enumerate}
\noindent In spite of these differences the two approaches seem to
agree within about a 20 \% accuracy.

As previously pointed out,
any model for the SRC SF, when integrated over the
removal energy in the momentum sum rule (Eq. (\ref{eq6-3a})), has
to provide the high momentum part
 of the one-nucleon momentum
distribution obtained by the Fourier transform of the non-diagonal one-nucleon density matrix produced
by the ground-state many-body wave functions. This is indeed the case of the convolution formula,
 as demonstrated
in Fig. \ref{Fig14}. Finally, in Fig. \ref{Fig15}, the convergence of the momentum sum rule
is shown: it can be seen that in order to correctly obtain the magnitude of the momentum distribution at $k \geq 4\,fm^{-1}$
the SF has to be integrated up to very high values of the removal energy ($E \simeq 400MeV$).

%%%%%%%%%%%%%%%%%%%%%%%%%%%%%%%%%%%%%%%%%%%%%%%%%%%%%%%%%%%%%%%%%%%%%%%%%%%%% S E C T I O N
\section{Summary and conclusion}
%%%%%%%%%%%%%%%%%%%%%%%%%%%%%%%%%%%%%%%%%%%%%%%%%%%%%%%%%%%%%%%%%%%%%%%%%%%%%%%%%%%%%%%%%%%
The main aspects and    results of the present  paper  can be listed as follows:
\begin{enumerate}

\item the NCLCE was used to minimize the nuclear Hamiltonian of light nuclei
containing
realistic model of the nucleon-nucleon interaction  and a comparison of the resulting binding energies, radii, one- and two-nucleon
momentum distributions, with particular emphasis on the high momentum components generated by SRC,
have been calculated and shown
to be in satisfactory agreement with the results of up-to-date approaches, like the VMC and CVMC ones;

\item we argued that  the basis of
any treatment of SRC  is wave function factorization at short range and, accordingly, by a detailed analysis of the
dependence of the two-nucleon momentum distribution $n_A^{N_1N_2}({\bf k}_{rel},{\bf K}_{c.m})$ upon
the relative, ${\bf k}_{rel}$, and c.m.,
${\bf K}_{c.m.}$, momenta of proton-neutron and proton-proton pairs embedded in the medium,  we have
demonstrated that  in the region of momenta governed by the short-range behavior of the NN interaction
($|{\bf k}_{rel}|\geq 1\, fm^{-1}$, $|{\bf k}_{rel}|>>|{\bf K}_{c.m}|$) the two-nucleon momentum
distributions factorize and the region
of factorization of the
nuclear wave function in momentum space has been clearly   identified;

\item  exploiting the factorization property of $n_A^{N_1N_2}({\bf k}_{rel},{\bf K}_{c.m})$ we
 have developed an advanced microscopic many-body, parameter-free approach
 to the SF which is expressed in terms of {\it ab-initio} A-dependent microscopic  relative and c.m. momentum distributions,
 reflecting
the underlying NN interaction; by this way,  the specific features of a given nucleus are taken into account
without recurring to any approximation;

\item in the case of the three-nucleon system, we found that the convolution formula
 fully agrees
 with the results of  the {\it ab-initio} SF in a wide interval of momenta and
 removal energy;

 \item in the case of complex nuclei the correctness of the convolution SF has been checked
 by means of the momentum sum rule, finding that the
 integral of the SF up to $E\simeq 400 MeV$ fully agrees up to $k\simeq 5 \, fm^{-1}$
 with  the exact one-nucleon momentum
 distribution, calculated independently  in terms of the ground-state wave functions.

\end{enumerate}

To summarize, we would like to stress that by exploiting the universal factorization property exhibited
 by the short-range behavior  of the
nuclear wave function
for finite nuclei, we have  generated a microscopic and parameter-free  SF
 based upon  a convolution of {\it ab initio} relative and c.m.
two-nucleon momentum distributions  for a given nucleus. The convolution SF rigorously satisfies the conditions
for its validity, in that  it takes
into account only those nucleon configurations compatible with the requirement
 of wave function factorization.  Our convolution approach for the three-nucleon systems provides results
 in full agreement with proton and neutron SF, whereas in complex nuclei,
  for which {\it ab-initio}
SF cannot yet be obtained, it fully satisfies the momentum sum rule.
These results,  coupled with the many-body microscopic nature of our approach and the absence of any
adjustable parameter, makes the convolution
SF a serious candidate for the investigation
 of nuclear effects in various processes, particularly in electro-weak scattering off nuclear targets.
 Needless to say
that these processes besides a realistic SF, also   require the inclusion
of all types of final-state interaction
which are at work when the active (struck) nucleon leaves the nucleus.
\newpage
\appendix
\section{The  one- and two-nucleon non diagonal density matrices with the NCLCE}\label{Appendix}
\subsection{One-nucleon non diagonal density matrix}
The one-nucleon non diagonal density matrix at first order of the NCLCE includes three terms, namely:
\begin{eqnarray}
  \label{One_nucleon_Total}
  \rho({\bf r}_1,{\bf r}_1')&=&\rho_{MF}({\bf r}_1,{\bf r}_{1}')+\rho_{2b}({\bf r}_1,{\bf r}_{1}')+\rho_{3b}({\bf r}_1,{\bf r}_{1}').
\end{eqnarray}
The suffixes (MF),(2b) and (3b) denote mean-field, 2-body and 3-body cluster term, respectively.
Each term of Eq. (\ref{One_nucleon_Total}) is expressed by using the density distributions in mean-field given by
\begin{eqnarray}
  \label{Rho_0}
  \rho_0({\bf r}_i)=\sum_{n,l,m}\left|\varphi_{nlm}({\bf r}_i) \right|^2, \ \
  \rho_0({\bf r}_i,{\bf r}_j)=\sum_{n,l,m}\varphi_{nlm}^*({\bf r}_i)\varphi_{nlm}({\bf r}_j) ,
\end{eqnarray}
where we take the following mean-field wave function
\begin{eqnarray}
  \label{Sell_Model}
  \psi_{MF}&=&\frac{1}{\sqrt{A!}}\textit{det}[\phi_{\alpha_i}(x_j)], \ \  \phi_{\alpha}(x_i)=\varphi_{nlm}({\bf r}_i)\chi(i)\zeta(i),
\end{eqnarray}
with $\chi(i)$ and $\zeta(i)$ being the spin and isospin wave function respectively.
The explicit form of each terms with the use of above quantities (Eq. (\ref{Rho_0})) is shown what follows.

\begin{flushleft}
\textit{\textbf{1.1 MF term}}
\begin{eqnarray}
  \label{One_Nucleon_Sell_Model}
  \rho_{SM}({\bf r}_1,{\bf r}_1')=4\rho_0({\bf r}_1,{\bf r}_1').
\end{eqnarray}
\end{flushleft}
\begin{flushleft}
\textit{\textbf{1.2 2-body term}}
\begin{eqnarray}
  \label{One_Nuclen_2b}
  \rho_{2b}({\bf r}_1,{\bf r}_1')=\frac{1}{A}\int d{\bf r}_2&\Bigl(&< 12|\hat{O}_{2b}|12>_{ST}\rho_0({\bf r}_1,{\bf r}_1')\rho_0({\bf r}_2) \\ \nonumber
  &-&  <12|\hat{O}_{2b}|21>_{ST}\rho_0({\bf r}_1,{\bf r}_2)\rho_0({\bf r}_2,{\bf r}_1')
   \Bigr),
\end{eqnarray}
where following definitions for the matrix elements in the spin-isospin space are introduced
\begin{eqnarray}
  \label{One_Nuclen_ST}
  < ij|\hat{O}_{2b}|kl>_{ST} &\equiv&
  <\chi(i)\chi(j)\zeta(i)\zeta(j)\,|\hat{O}_{2b}|\,\chi(k)\chi(l)\zeta(k)\zeta(l)>, \\ \nonumber
  \hat{O}_{2b}&\equiv& \hat{f}(12)\hat{f}(1'2)-1.
\end{eqnarray}
\end{flushleft}
\begin{flushleft}
\textit{\textbf{1.3 3-body term}}
\begin{eqnarray}
  \label{One_Nuclen_3b}
  \rho_{3b}({\bf r}_1,{\bf r}_1')&=&\frac{1}{A}\int d{\bf r}_2 d{\bf r}_3\rho_0({\bf r}_1,{\bf r}_2) \\  \nonumber
  &\times&\left(< 123|\hat{O}_{3b}|231>_{ST}\rho_0({\bf r}_2,{\bf r}_3)\rho_0({\bf r}_3,{\bf r}_1')
  -  <123|\hat{O}_{3b}|213>_{ST}\rho_0({\bf r}_2,{\bf r}_1')\rho_0({\bf r}_3)
   \right),\\
   \hat{O}_{3b}&\equiv& \hat{f}(23)\hat{f}(23)-1 .
\end{eqnarray}
\end{flushleft}

\subsection{Two-nucleon non diagonal density matrix}
The two-nucleon non diagonal density matrix at first order of the NCLCE includes four terms, as follows
\begin{eqnarray}
  \label{Two_nucleon_Total}
  \rho^{pN}({\bf r}_1,{\bf r}_2,{\bf r}_1',{\bf r}_2')&=&\rho_{MF}^{pN}({\bf r}_1,{\bf r}_2,{\bf r}_1',{\bf r}_2')
  +\rho_{2b}^{pN}({\bf r}_1,{\bf r}_2,{\bf r}_1',{\bf r}_2')
  +\rho_{3b}^{pN}({\bf r}_1,{\bf r}_2,{\bf r}_1',{\bf r}_2')  \\ \nonumber
  &+&\rho_{4b}^{pN}({\bf r}_1,{\bf r}_2,{\bf r}_1',{\bf r}_2') .
\end{eqnarray}
The explicit forms of each term in Eq. (\ref{Two_nucleon_Total}) are summarized what follows.
\begin{flushleft}
\textit{\textbf{2.1 MF term}}
  \begin{eqnarray}
    \rho_{MF}^{pn}({\bf r}_1,{\bf r}_2,{\bf r}_1',{\bf r}_2')&=&\frac{1}{A(A-1)}8\rho_0({\bf r}_1,{\bf r}_1')\rho_0({\bf r}_2,{\bf r}_2'), \\
    \rho_{MF}^{pp}({\bf r}_1,{\bf r}_2,{\bf r}_1',{\bf r}_2')&=&\frac{2}{A(A-1)}(2\rho_0({\bf r}_1,{\bf r}_1')\rho_0({\bf r}_2,{\bf r}_2')-
    \rho_0({\bf r}_1,{\bf r}_2')\rho_0({\bf r}_{2},{\bf r}_1') ) .
    \label{Two_nucleon_SM}
  \end{eqnarray}
  \end{flushleft}
\begin{flushleft}
\textit{\textbf{2.2 2-body term}}
  \begin{eqnarray}
    \rho_{2b}^{pN}({\bf r}_1,{\bf r}_2,{\bf r}_1',{\bf r}_2')&=&\frac{1}{A(A-1)}\Bigl(<12|\hat{O}_{2b}|12>_{ST}\rho_0({\bf r}_1,{\bf r}_1')\rho_0({\bf r}_2,{\bf r}_2') \\ \nonumber
    &&-<12|\hat{O}_{2b}|21>_{ST}\rho_0({\bf r}_1,{\bf r}_{2}) \rho_0({\bf r}_2,{\bf r}_1')\Bigr) ,\\
   \hat{O}_{2b}&\equiv& \left(\hat{f}(12)\hat{f}(1'2')-1\right)\hat{P}^{pN}(12) ,
  \label{Two_nucleon_2b}
  \end{eqnarray}
    where $\hat{P}^{pN}(ij)$ is a projection operator on the pN pair.
  \end{flushleft}
\begin{flushleft}
\textit{\textbf{2.2 3-body term}}
   \begin{eqnarray}
    \rho_{3b}^{pN}({\bf r}_1,{\bf r}_2,{\bf r}_1',{\bf r}_2')&=&\frac{2}{A(A-1)}
    \int d{\bf r}_3 \textit{F}_{3b}^{pN}({\bf r}_1,{\bf r}_2,{\bf r}_1',{\bf r}_2',{\bf r}_3),\\
   \textit{F}_{3b}^{pN}&=&\langle 123|\hat{O}_{3b}|123\rangle_{ST}\rho_0({\bf r}_1,{\bf r}_{1'})\rho_0({\bf r}_2,{\bf r}_{2'})\rho_0({\bf r}_3) \\ \nonumber
  &+&\langle 123|\hat{O}_{3b}|231\rangle_{ST}\rho_0({\bf r}_1,{\bf r}_{2'})\rho_0({\bf r}_2,{\bf r}_{3})\rho_0({\bf r}_3,{\bf r}_{1'}) \\ \nonumber
  &+&\langle 123|\hat{O}_{3b}|312\rangle_{ST}\rho_0({\bf r}_1,{\bf r}_{3})\rho_0({\bf r}_2,{\bf r}_{1'})\rho_0({\bf r}_3,{\bf r}_{2'}) \\ \nonumber
  &-&\langle 123|\hat{O}_{3b}|132\rangle_{ST}\rho_0({\bf r}_1,{\bf r}_{1'})\rho_0({\bf r}_2,{\bf r}_{3})\rho_0({\bf r}_3,{\bf r}_{2'}) \\ \nonumber
  &-&\langle 123|\hat{O}_{3b}|213\rangle_{ST}\rho_0({\bf r}_1,{\bf r}_{2'})\rho_0({\bf r}_2,{\bf r}_{1'})\rho_0({\bf r}_{3}) \\ \nonumber
  &-&\langle 123|\hat{O}_{3b}|321\rangle_{ST}\rho_0({\bf r}_1,{\bf r}_{3})\rho_0({\bf r}_2,{\bf r}_{2'})\rho_0({\bf r}_3,{\bf r}_{1'}) , \\
  \hat{O}_{3b}&\equiv& \left( \hat{f}(13)\hat{f}(1'3)-1\right)\hat{P}^{pN}(12) .
    \label{Two_nucleon_3b}
    \end{eqnarray}
   \end{flushleft}
\begin{flushleft}
\textit{\textbf{2.4 4-body term}}
    \begin{eqnarray}
      \rho_{4b}^{pN}&(&{\bf r}_1,{\bf r}_2,{\bf r}_1',{\bf r}_2')=\frac{1}{2A(A-1)}
    \int d{\bf r}_3 d{\bf r}_4 \textit{F}_{4b}^{pN}({\bf r}_1,{\bf r}_2,{\bf r}_1',{\bf r}_2',{\bf r}_3,{\bf r}_4),\\
    \textit{F}_{4b}^{pN}&=&
      <1234|\hat{O}_{4b}|2314>_{ST}\bigl(
        \rho_0({\bf r}_1,{\bf r}_2')\rho_0({\bf r}_2,{\bf r}_{3})\rho_0({\bf r}_3,{\bf r}_1')\rho_0({\bf r}_4) \\ \nonumber
        &+& \rho_0({\bf r}_1,{\bf r}_{2'})\rho_0({\bf r}_2,{\bf r}_{4})\rho_0({\bf r}_3)\rho_0({\bf r}_4,{\bf r}_{1'})
         +\rho_0({\bf r}_1,{\bf r}_{3})\rho_0({\bf r}_2,{\bf r}_{1'})\rho_0({\bf r}_3,{\bf r}_{2'})\rho_0({\bf r}_4) \\ \nonumber
         &+&\rho_0({\bf r}_1,{\bf r}_{4})\rho_0({\bf r}_2,{\bf r}_{1'})\rho_0({\bf r}_3)\rho_0({\bf r}_4,{\bf r}_{2'})
         \bigr)\\ \nonumber
        &+&<1234|\hat{O}_{4b}|1342>_{ST}\bigl(
         \rho_0({\bf r}_1,{\bf r}_1')\rho_0({\bf r}_2,{\bf r}_{3})\rho_0({\bf r}_3,{\bf r}_{4})\rho_0({\bf r}_4,{\bf r}_2') \\ \nonumber
        &+& \rho_0({\bf r}_1,{\bf r}_1')\rho_0({\bf r}_2,{\bf r}_{4})\rho_0({\bf r}_3,{\bf r}_2')\rho_0({\bf r}_4,{\bf r}_{3})
          +\rho_0({\bf r}_1,{\bf r}_{3})\rho_0({\bf r}_2,{\bf r}_2')\rho_0({\bf r}_3,{\bf r}_{4})\rho_0({\bf r}_4,{\bf r}_1') \\ \nonumber
          &+&\rho_0({\bf r}_1,{\bf r}_{4})\rho_0({\bf r}_2,{\bf r}_2')\rho_0({\bf r}_3,{\bf r}_1')\rho_0({\bf r}_4,{\bf r}_{3})
        \bigr)\\ \nonumber
        &+&<1234|\hat{O}_{4b}|3412>_{ST}\bigl(
          \rho_0({\bf r}_1,{\bf r}_{3})\rho_0({\bf r}_2,{\bf r}_{4})\rho_0({\bf r}_3,{\bf r}_1')\rho_0({\bf r}_4,{\bf r}_2') \\ \nonumber
         &+& \rho_0({\bf r}_1,{\bf r}_{4})\rho_0({\bf r}_2,{\bf r}_{3})\rho_0({\bf r}_3,{\bf r}_2')\rho_0({\bf r}_4,{\bf r}_1')
         \bigr)\\ \nonumber
         &-&<1234|\hat{O}_{4b}|1324>_{ST}\bigl(
           \rho_0({\bf r}_1,{\bf r}_1')\rho_0({\bf r}_2,{\bf r}_{3})\rho_0({\bf r}_3,{\bf r}_2')\rho_0({\bf r}_4) \\ \nonumber
         &+& \rho_0({\bf r}_1,{\bf r}_1')\rho_0({\bf r}_2,{\bf r}_{4})\rho_0({\bf r}_3)\rho_0({\bf r}_4,{\bf r}_2')
          +\rho_0({\bf r}_1,{\bf r}_{3})\rho_0({\bf r}_2,{\bf r}_2')\rho_0({\bf r}_3,{\bf r}_1')\rho_0({\bf r}_4) \\ \nonumber
         &+&\rho_0({\bf r}_1,{\bf r}_{4})\rho_0({\bf r}_2,{\bf r}_2')\rho_0({\bf r}_3)\rho_0({\bf r}_4,{\bf r}_1')
         \bigr)\\ \nonumber
         &-&<1234|\hat{O}_{4b}|2341>_{ST}\bigl(
           \rho_0({\bf r}_1,{\bf r}_2')\rho_0({\bf r}_2,{\bf r}_{3})\rho_0({\bf r}_3,{\bf r}_{4})\rho_0({\bf r}_4,{\bf r}_1') \\ \nonumber
         &+& \rho_0({\bf r}_1,{\bf r}_2')\rho_0({\bf r}_2,{\bf r}_{4})\rho_0({\bf r}_3,{\bf r}_1')\rho_0({\bf r}_4,{\bf r}_{3})
             +\rho_0({\bf r}_1,{\bf r}_{3})\rho_0({\bf r}_2,{\bf r}_1')\rho_0({\bf r}_3,{\bf r}_{4})\rho_0({\bf r}_4,{\bf r}_2') \\ \nonumber
          &+&\rho_0({\bf r}_1,{\bf r}_{4})\rho_0({\bf r}_2,{\bf r}_1')\rho_0({\bf r}_3,{\bf r}_2')\rho_0({\bf r}_4,{\bf r}_{3})
         \bigr)\\ \nonumber
         &-&<1234|\hat{O}_{4b}|3421>_{ST}\bigl(
           \rho_0({\bf r}_1,{\bf r}_{3})\rho_0({\bf r}_2,{\bf r}_{4})\rho_0({\bf r}_3,{\bf r}_2')\rho_0({\bf r}_4,{\bf r}_1') \\ \nonumber
         &+& \rho_0({\bf r}_1,{\bf r}_{4})\rho_0({\bf r}_2,{\bf r}_{3})\rho_0({\bf r}_3,{\bf r}_1')\rho_0({\bf r}_4,{\bf r}_2')
       \bigr) , \\
      \hat{O}_{4b}&\equiv& \left( \hat{f}(34)\hat{f}(34)-1\right)\hat{P}^{pN}(12) .
    \label{Two_nucleon_4b}
    \end{eqnarray}
   \end{flushleft}

\newpage
%%%%%%%%%%%%%%%%%%%%%%%%%%%%%%%%%%%%%%%%%%%%%%%%%%%%%%%%%%%%%%%%%%%%%%%% N E W P A G E

%%%%%%%%%%%%%%%%%%%%%%%%%%%%%%%%%%%%%%%%%%%%%%%%%%%%%%%%%%%%%%%%%%%%%%%%%%%%%%%

  %%%%%%%%%%%%%%%%%%%%%%%%%%%%%%%%%%%%%%%%%%%%%%%%%%%%%%%%%%%%%%%%%%%%%%%%%%%%%%%%%%%%%%%%%%%%%% F I G U R E S
\newpage

    \begin{table}[!hp]
   \caption{A comparison of the results of three-realistic  many-body calculations
    for the ground-state energy and r.m.s.
   radius of $A=16$ obtained by the minimization of Eq. (\ref{2:EV}):
 Cluster Variational Monte Carlo (CVMC) \cite{CVMC_ARGO}, Normalization
 Conserving Linked
Cluster Expansion (NCLCE)
    \cite{2:Alvioli:2005cz}. The three methods are variational ones and use Woods-Saxon
   single-particle wave functions and  similar
    2N interactions, with and without 3N $UIX$ interaction.
    Energies in $MeV$ and radii in $fm$.}
     \begin{center}
      \begin{tabular}{l||c|c|c|c|c|c}
Mean Field &Approach&$ Potential$&$
(E/A)$&$(E/A)_{exp}$&${<r^2>}^{1/2}$&$({<r^2>}^{1/2})_{exp}$\\\hline
WS& NCLCE
& AV8' & -4.4 & -7.98 & 2.64& 2.69\\
WS&\textit{CVMC}
                 & AV18 & -5.5 & -7.98 & 2.54& 2.69\\
                 WS&\textit{CVMC}
                 & AV18+UIX& -5.15 & -7.98 & 2.74& 2.69\\\hline
      \end{tabular}
      \end{center}
      %\end{center}
\label{Table1}
\end{table}
%%%%%%%%%%%%%%%%%%%%%%%%%%%%%%%%%%%%%%%%%%%%%%%%%%%%%%%%%%%%%%%%%%%%%%%% N E W P A G E
\newpage
%%%%%%%%%%%%%%%%%%%%%%%%%%%%%%%%%%%%%%%%%%%%%%%%%%%%%%%%%%%%%%%%%%%%%%%% N E W F I G U R E
\begin{figure}
\centerline{
\includegraphics[scale=0.8]{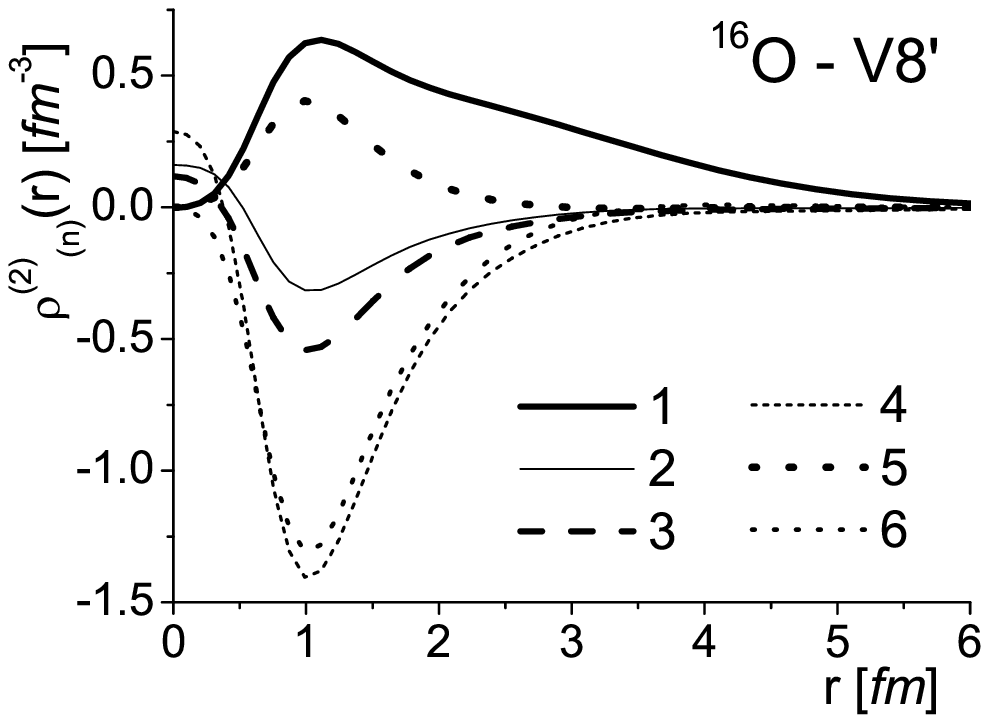}
\includegraphics[scale=0.8]{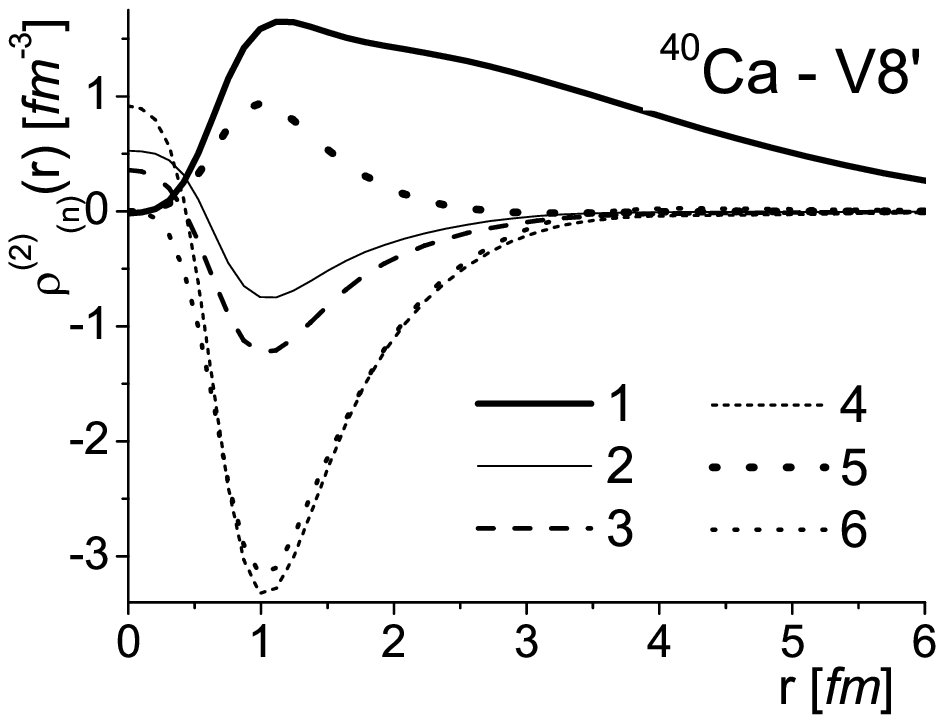}}
 \vspace{1.5cm}
\centerline{\includegraphics[scale=0.3]{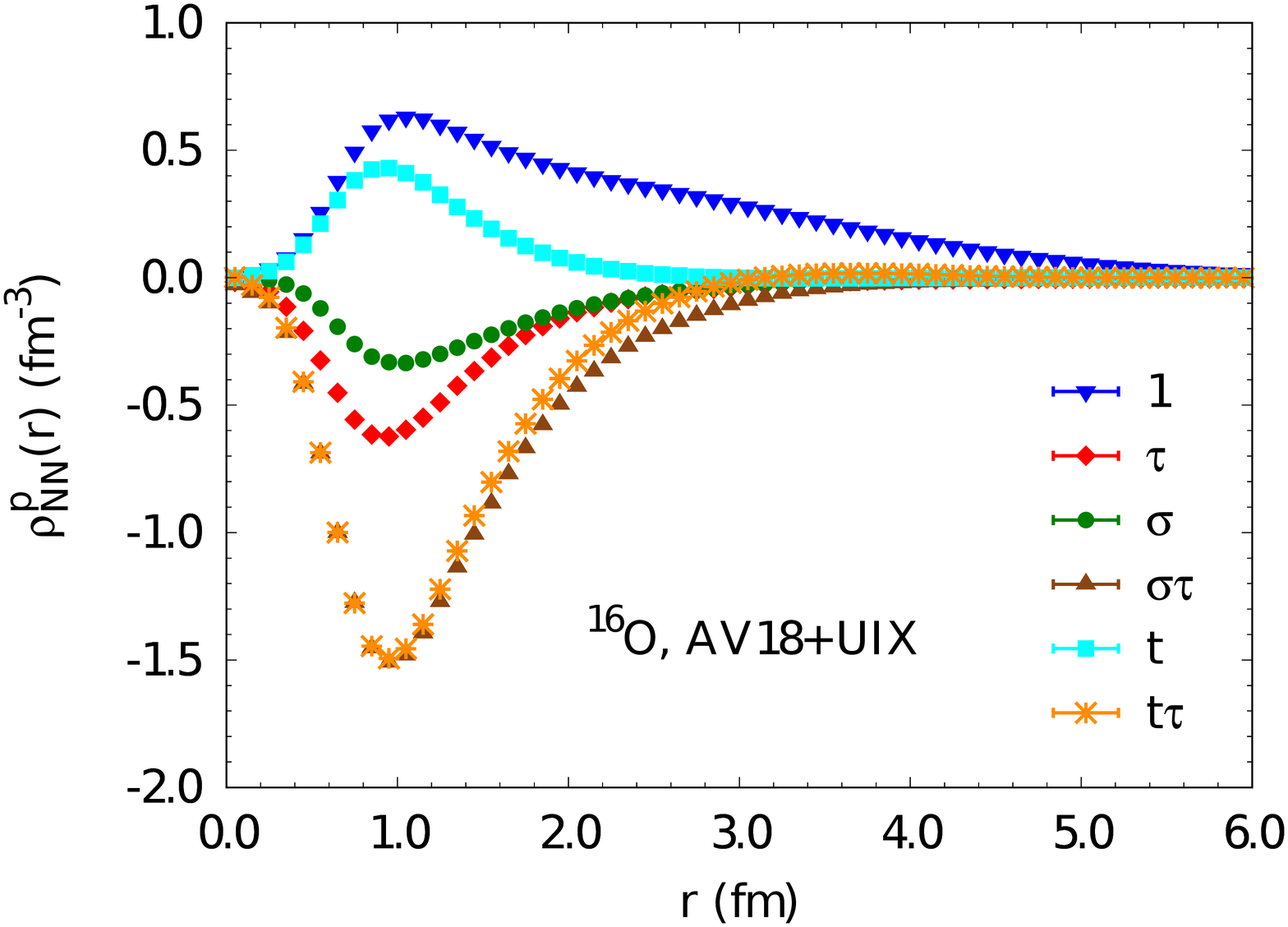}
\includegraphics[scale=0.3]{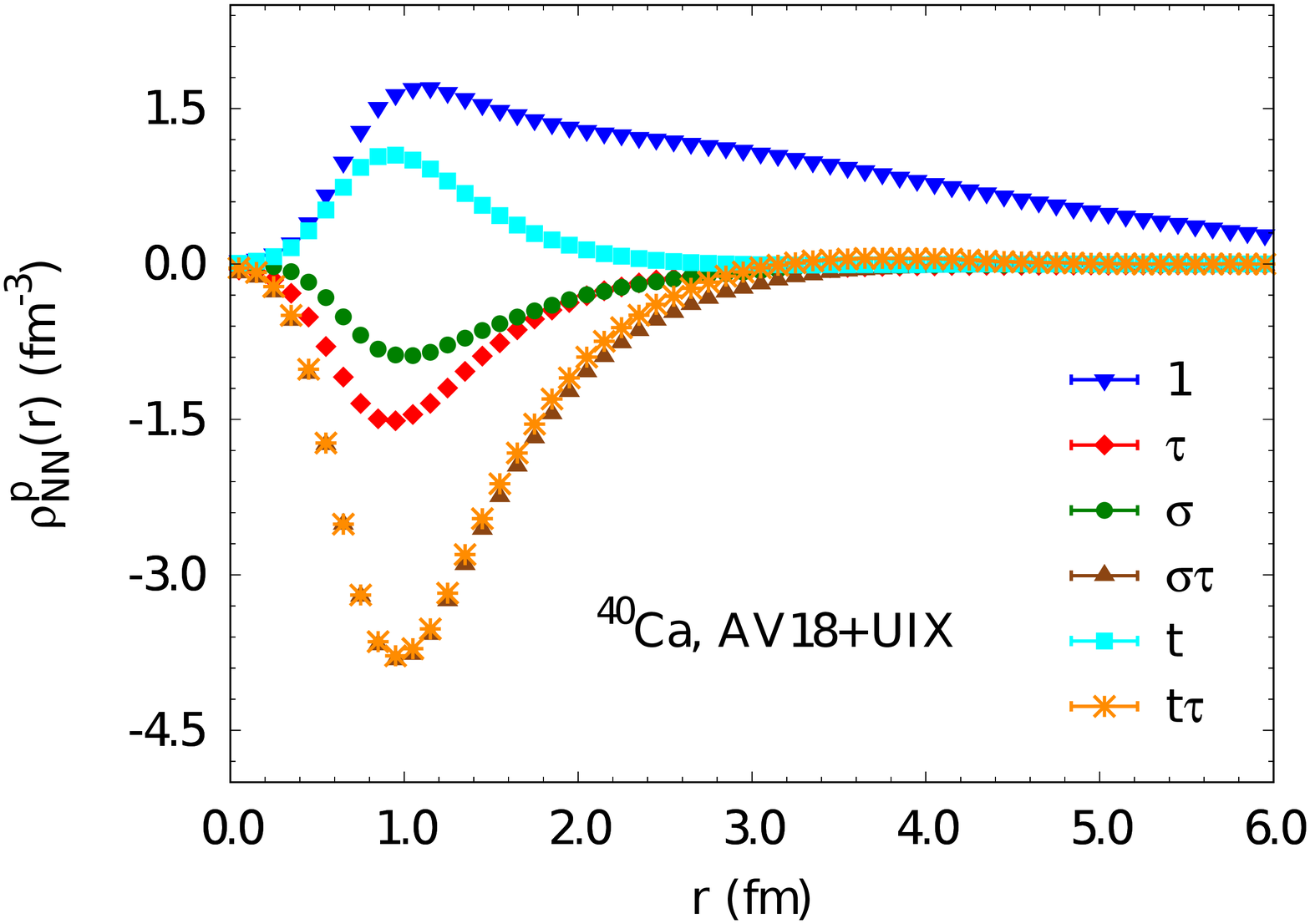}}
\caption{{\bf Upper panel}: the two-body density
$\rho_n^{(2)}(r=|{\bf r}_1- {\bf r}_2|)$  obtained in the
variational NCLCE calculation of Ref. \cite{2:Alvioli:2005cz}
performed with the first six components  of the Argonne
$V8^{\prime}$ $NN$ interaction (Eq. (\ref{n2_rel})) corresponding to
the values of the ground-state energy and radius  listed in Table
\ref{Table1}. {\bf Lower panel}: the same as in the upper panel
but in the case of
 the calculation  of Ref. \cite{CVMC_ARGO} performed with the $AV18$ $NN$
  interaction plus  $UIX$ $3N$ interaction.}
 \label{Fig1}
\end{figure}
\newpage
%%%%%%%%%%%%%%%%%%%%%%%%%%%%%%%%%%%%%%%%%%%%%%%%%%%%%%%%%%%%%%%%%%%%%%%% N E W F I G U R E
\begin{figure}
\centerline{\includegraphics[scale=0.35]{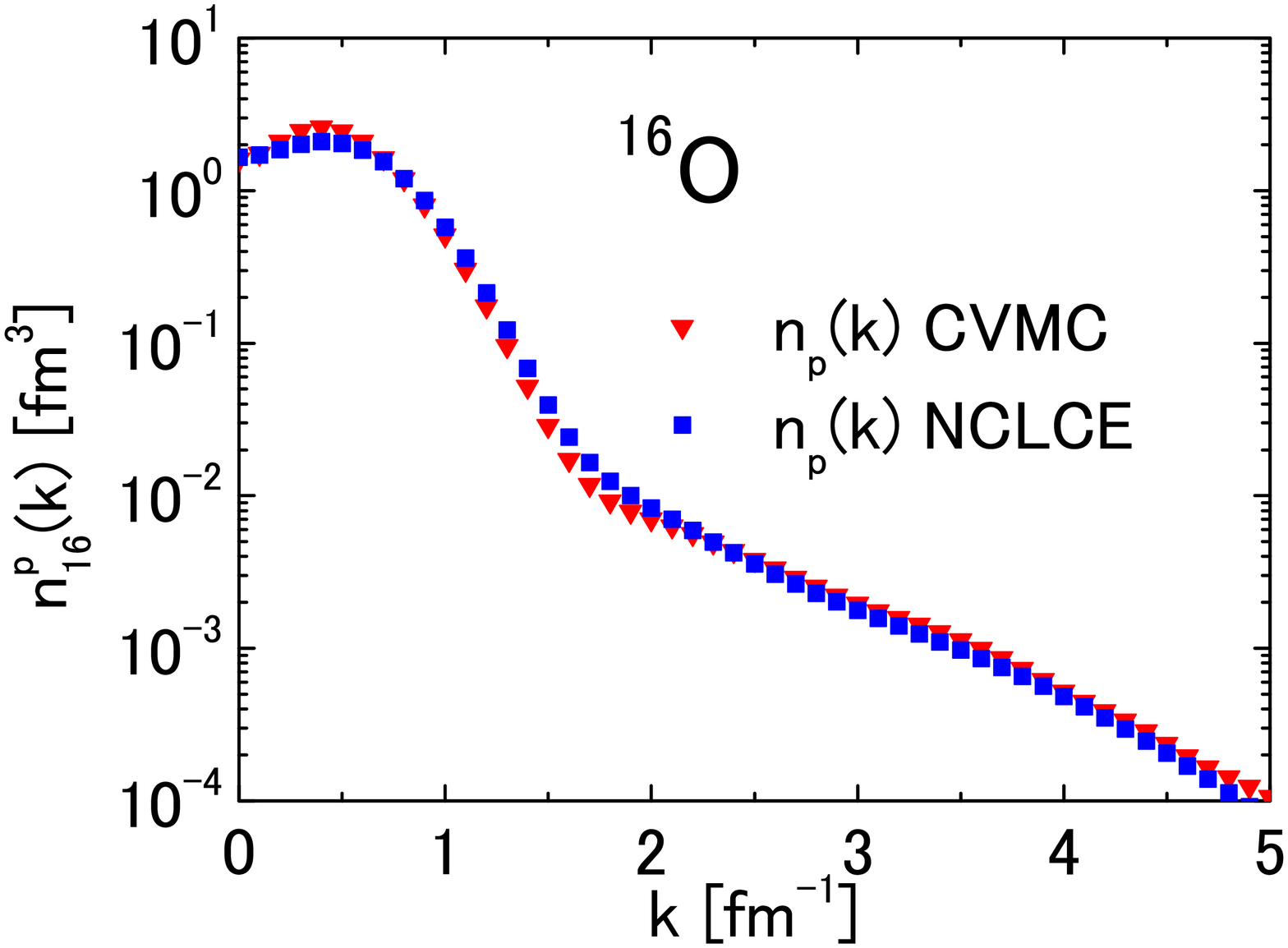}}
  %\hspace{-1.5cm}
  \centerline{
\includegraphics[scale=0.35]{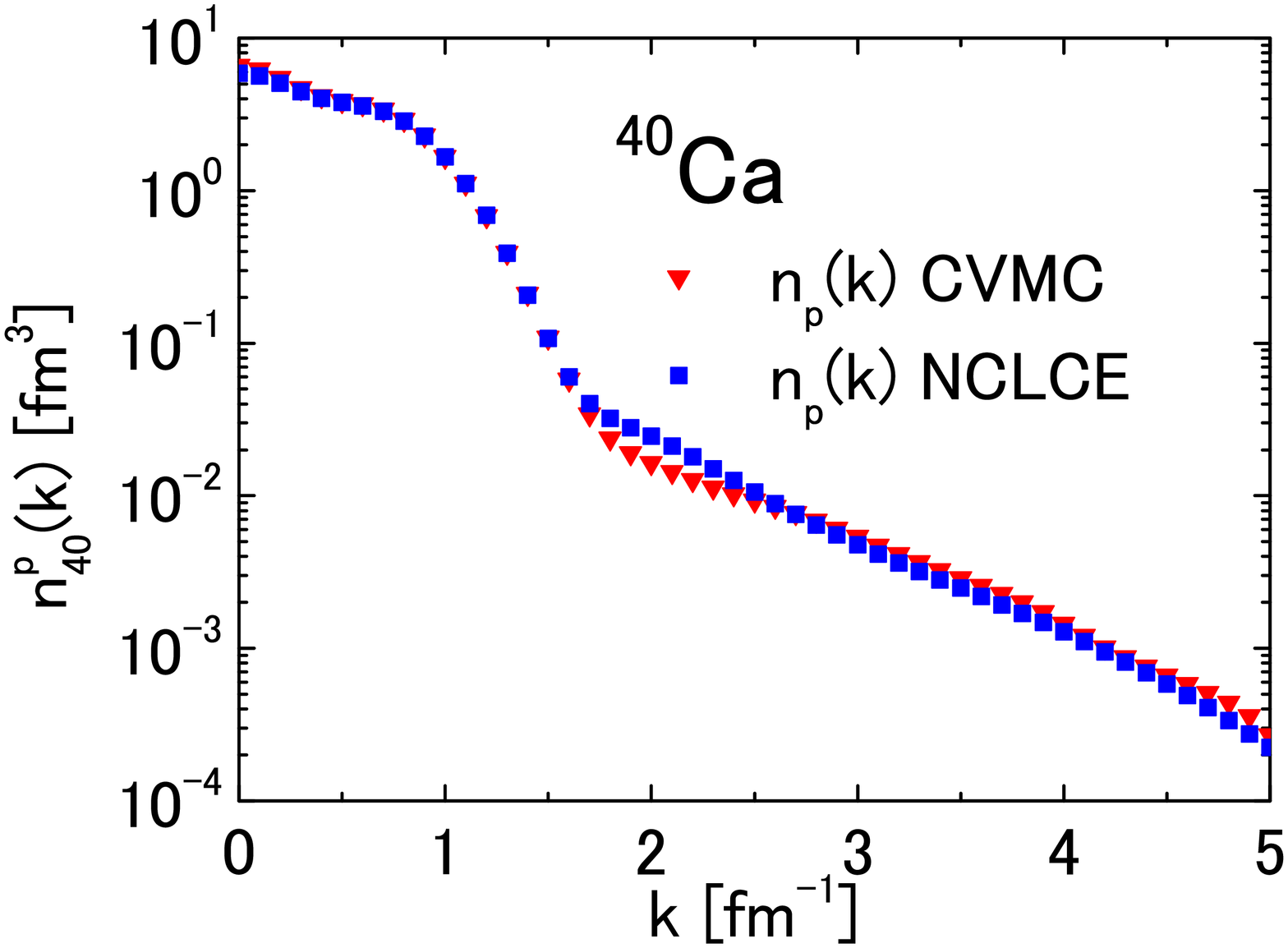}}
\caption{The proton ($n_p=n_n$) one-nucleon momentum distribution
of $^4He$ and $^{16}O$ obtained in Ref. \cite{CVMC_ARGO} within
the Cluster Variational Monte Carlo (CVMC)  and in Ref.
\cite{2:Alvioli:2005cz} within the NCLCE  at the lowest order.}
 \label{Fig2}
\end{figure}

\newpage
%%%%%%%%%%%%%%%%%%%%%%%%%%%%%%%%%%%%%%%%%%%%%%%%%%%%%%%%%%%%%%%%%%%%%%%% N E W F I G U R E
\begin{figure}[ht]
\begin{center}
\includegraphics[width=0.85\textwidth,keepaspectratio]{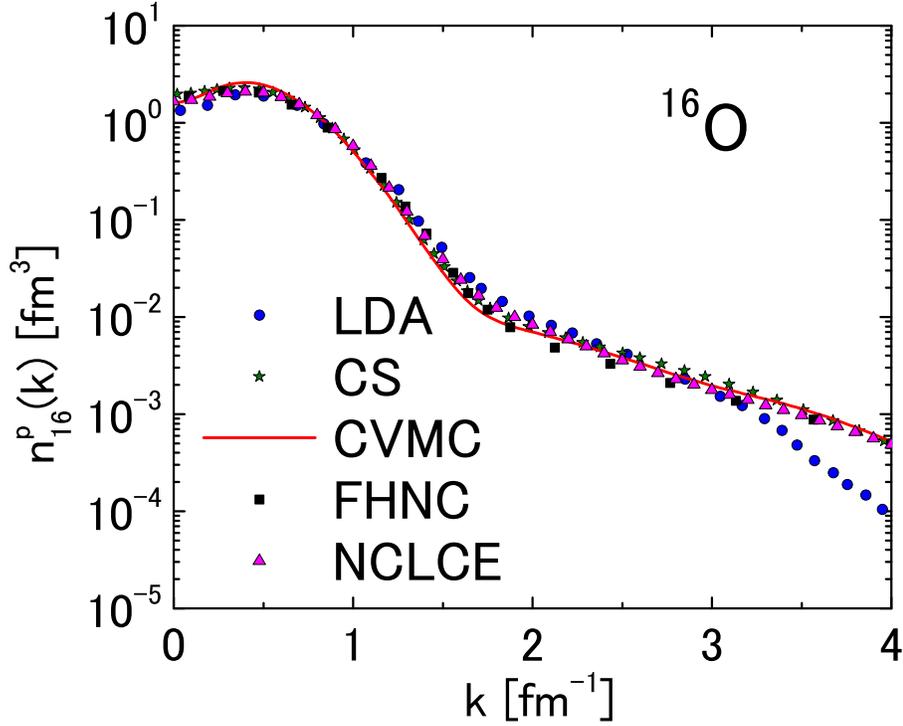}
\caption{(Color online) The  momentum distributions
   of $^{16}$O calculated with different many-body approaches
   and similar realistic $NN$ interactions:
    the Cluster Variational Monte Carlo (CVMC) results of Ref. \cite{CVMC_ARGO} (full line);
    the Normalization Conserving Linked Cluster Expansion
    (NCLCE)
   Ref. \cite{2:Alvioli:2005cz} (triangles); the
    Fermion  Hypernetted Chain Method (FHNC)
   of Ref. \cite{AriasdeSaavedra:2007byz}   with V8$^{\prime}$ interaction (squares);
   the
    two-nucleon correlation model (CS)
   of Ref. \cite{CiofidegliAtti:1995qe} (asterisks); the full
   dots represent the momentum distributions  obtained
  by integrating the SF obtained   within the
 nuclear matter Local Density Approximation
 (LDA) \cite{Omar_PkE,Omar_Neutrino}.}
\label{Fig3}
\end{center}
\end{figure}
\newpage
%%%%%%%%%%%%%%%%%%%%%%%%%%%%%%%%%%%%%%%%%%%%%%%%%%%%%%%%%%%%%%%%%%%%% N E W F I G U R E
\begin{figure}[ht]
\begin{center}
\includegraphics[width=0.85\textwidth,keepaspectratio]{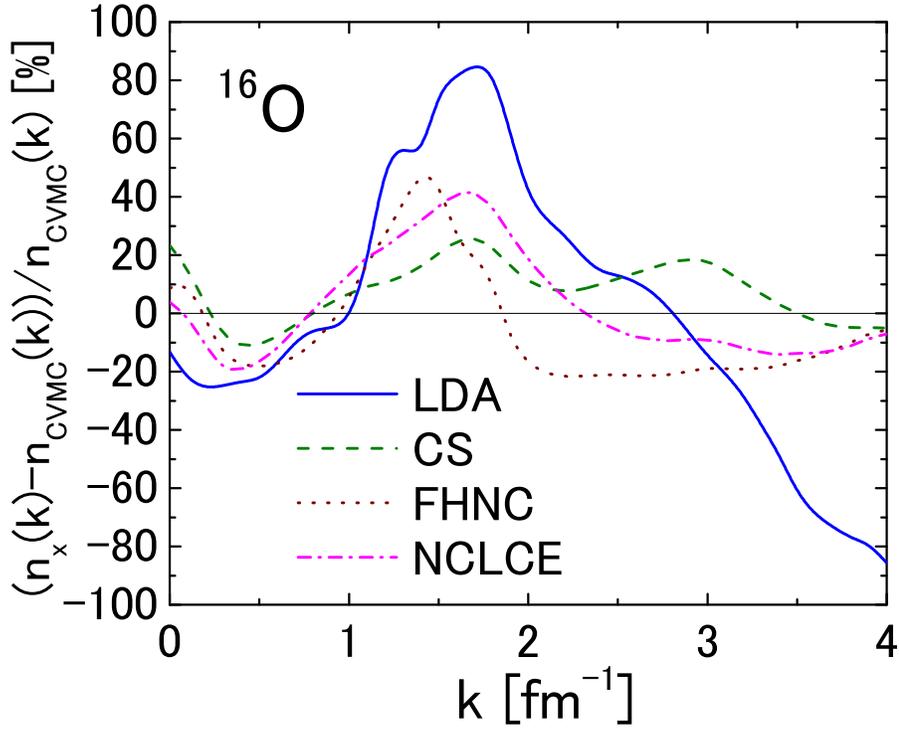}
\caption{(Color online)  The quantity
$100\frac{\Delta\,n(k)}{n^{VMCV}}$ (Eq. (\ref{deviation})) i.e.  the
percent deviation of  the   microscopic calculations of the
momentum distributions
   of $^{16}$O shown in
Fig. \ref{Fig3} taking the CVMC results of  Ref. \cite{CVMC_ARGO}
as the reference results.
    LDA: Ref. \cite{Omar_PkE,Omar_Neutrino};  CS: Ref. \cite{CiofidegliAtti:1995qe};
    FHNC: Ref. \cite{AriasdeSaavedra:2007byz}; NCLCE:
    Ref. \cite {2:Alvioli:2005cz}.}
    \label{Fig4}
\end{center}
\end{figure}
\newpage
%%%%%%%%%%%%%%%%%%%%%%%%%%%%%%%%%%%%%%%%%%%%%%%%%%%%%%%%%%%%%%%%%%%%%%%%%% N E W F I G U R E
\begin{figure}[ht]
\begin{center}
\includegraphics[width=1.0\textwidth,keepaspectratio]{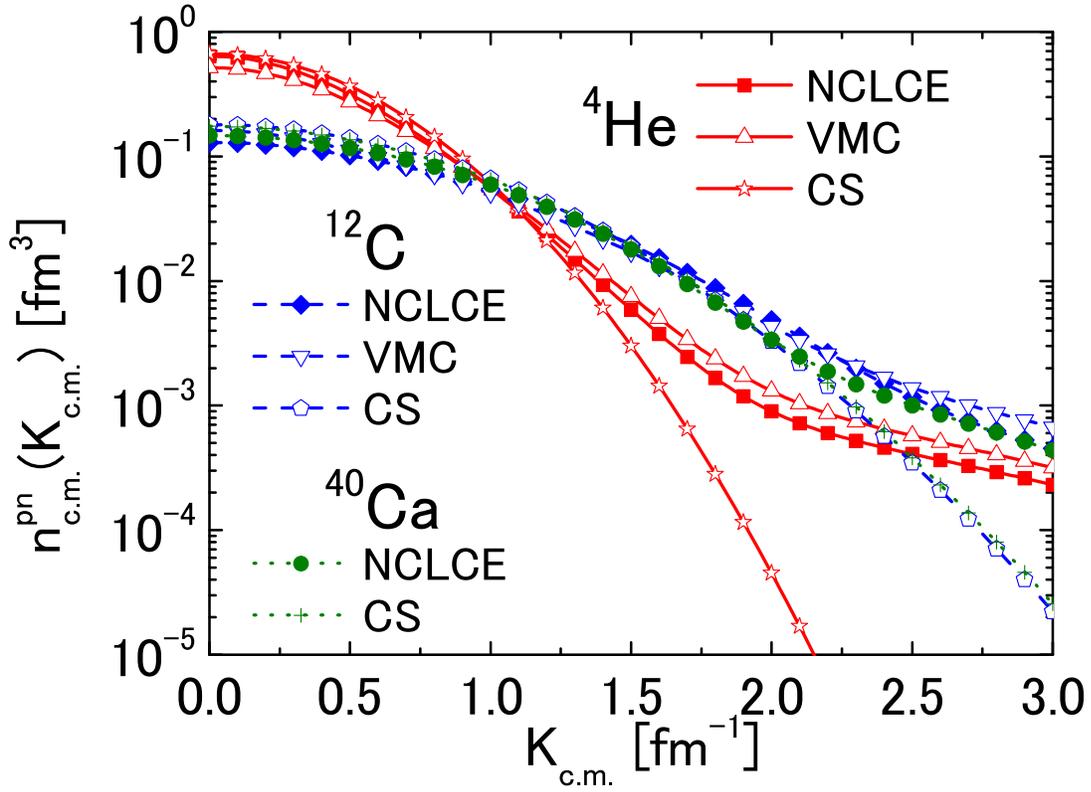}
 \vskip -0.5cm
 \caption{(Color online)
The proton-neutron center-of-mass (c.m.) momentum distributions
 (Eq. (\ref{n2_CM})) in $^{4}$He,  $^{12}$C and $^{40}$Ca calculated within
  microscopic many-body approaches. NCLCE: Ref.
  \cite{2:Alvioli:2016}; VMC: Ref.
  \cite{Wiringa:2013ala}; CS: Ref. \cite{CiofidegliAtti:1995qe}.}
 \label{Fig5}
\end{center}
\end{figure}
%%%%%%%%%%%%%%%%%%%%%%%%%%%%%%%%%%%%%%%%%%%%%%%%%%%%%%%%%%%%%%%%%%%%%%%%%% N E W F I G U R E
\newpage
\begin{figure}[ht]
\begin{center}
\includegraphics[width=0.9\textwidth,keepaspectratio]{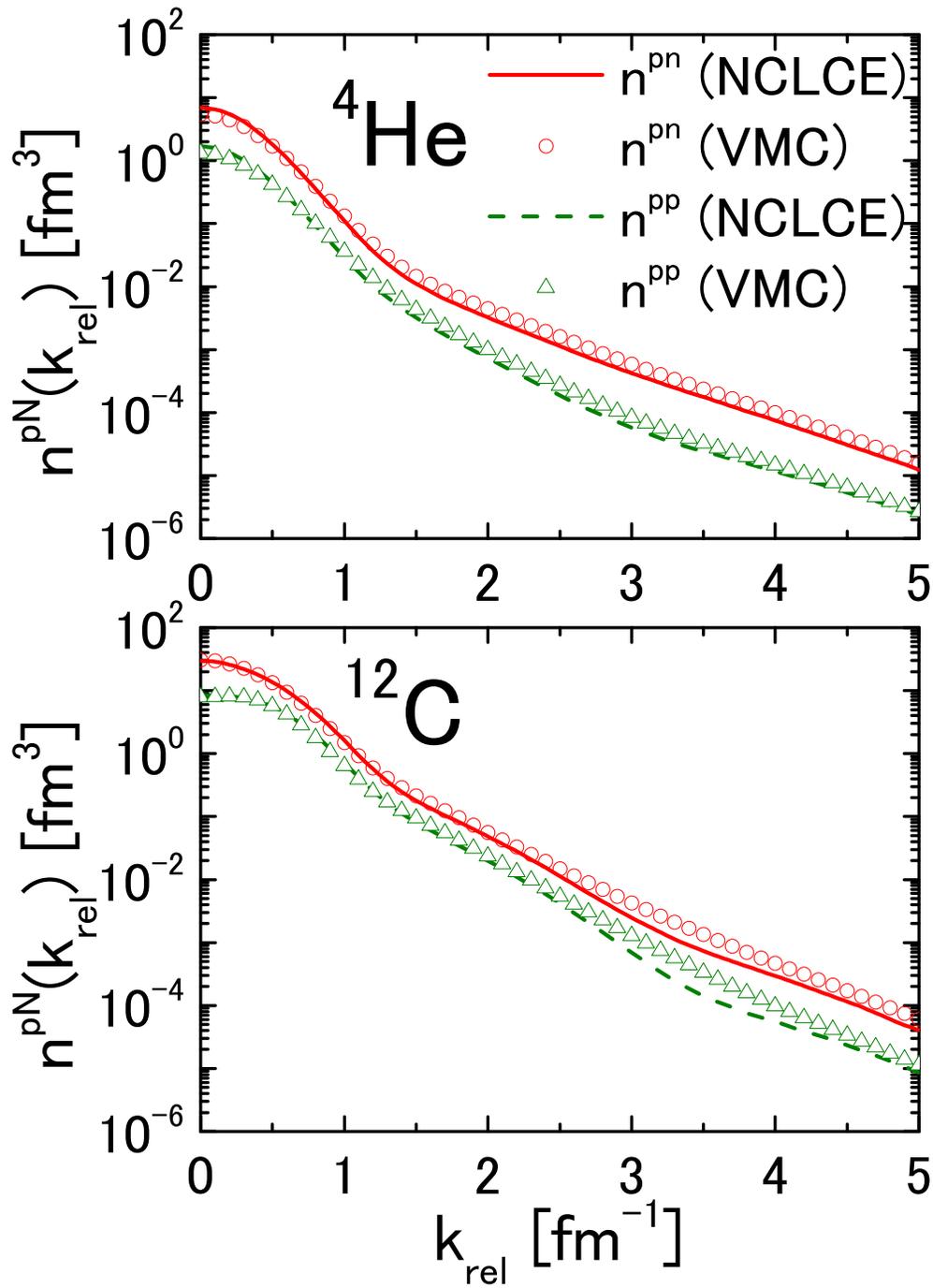}
 \vskip -0.5cm
 \caption{(Color online)
The $pn$ and  $pp$ relative momentum distributions (Eq.
(\ref{n2_rel_2})) in $^4$He and $^{12}C$ calculated by the NCLCE
(lines) \cite{2:Alvioli:2016} and the VMC
(symbols) \cite {Wiringa:2013ala}.}
 \label{Fig6}
\end{center}
\end{figure}
%%%%%%%%%%%%%%%%%%%%%%%%%%%%%%%%%%%%%%%%%%%%%%%%%%%%%%%%%%%%%%%%%%%%%%%%%% N E W F I G U R E
\newpage
\begin{figure}
%\begin{center}
\centerline{
\includegraphics[width=0.7\textwidth,keepaspectratio]
{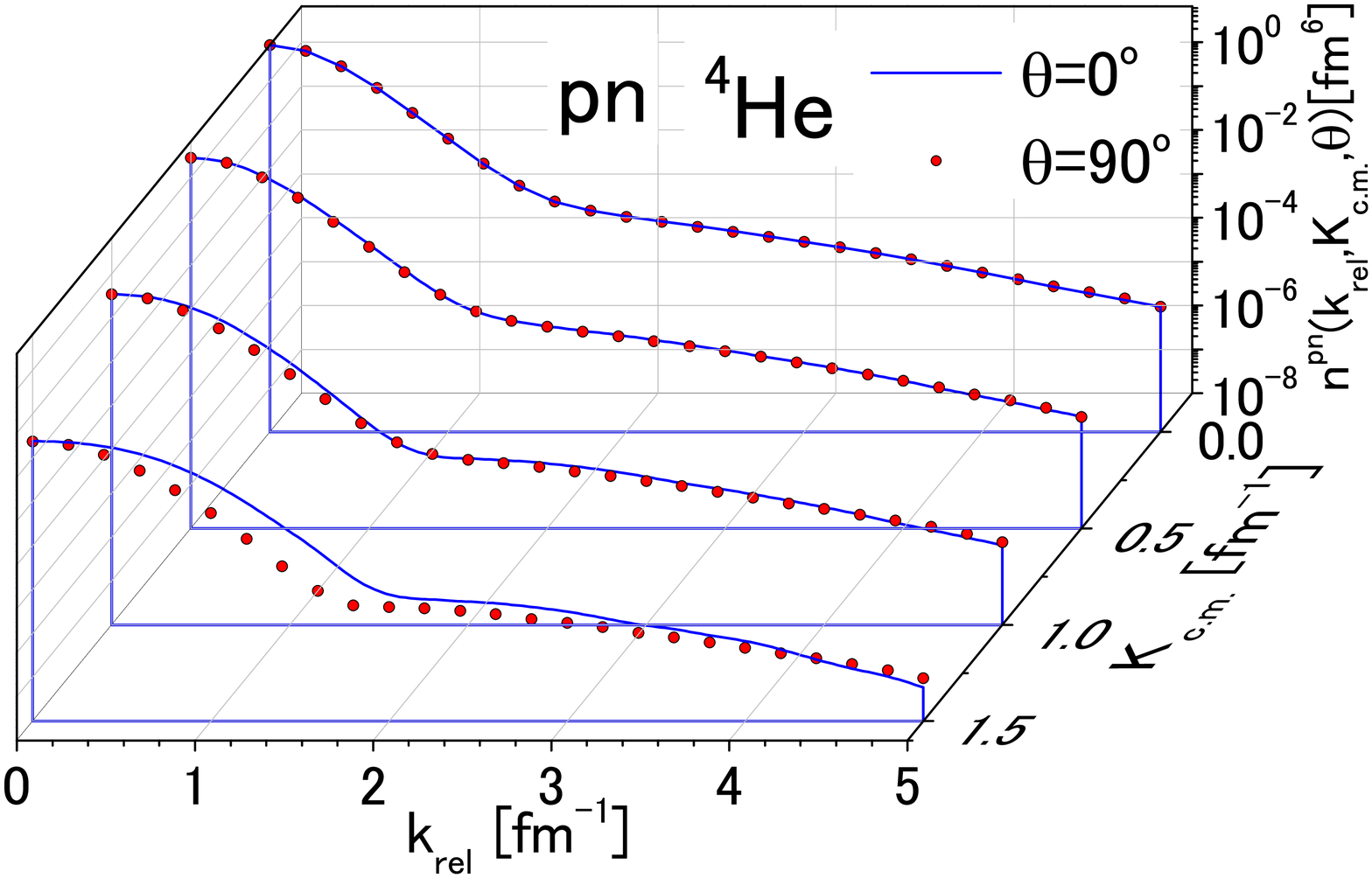} }
 \vskip -1.0cm %\hskip -2.0cm
 \centerline{
 \includegraphics[width=0.7\textwidth,keepaspectratio]
 {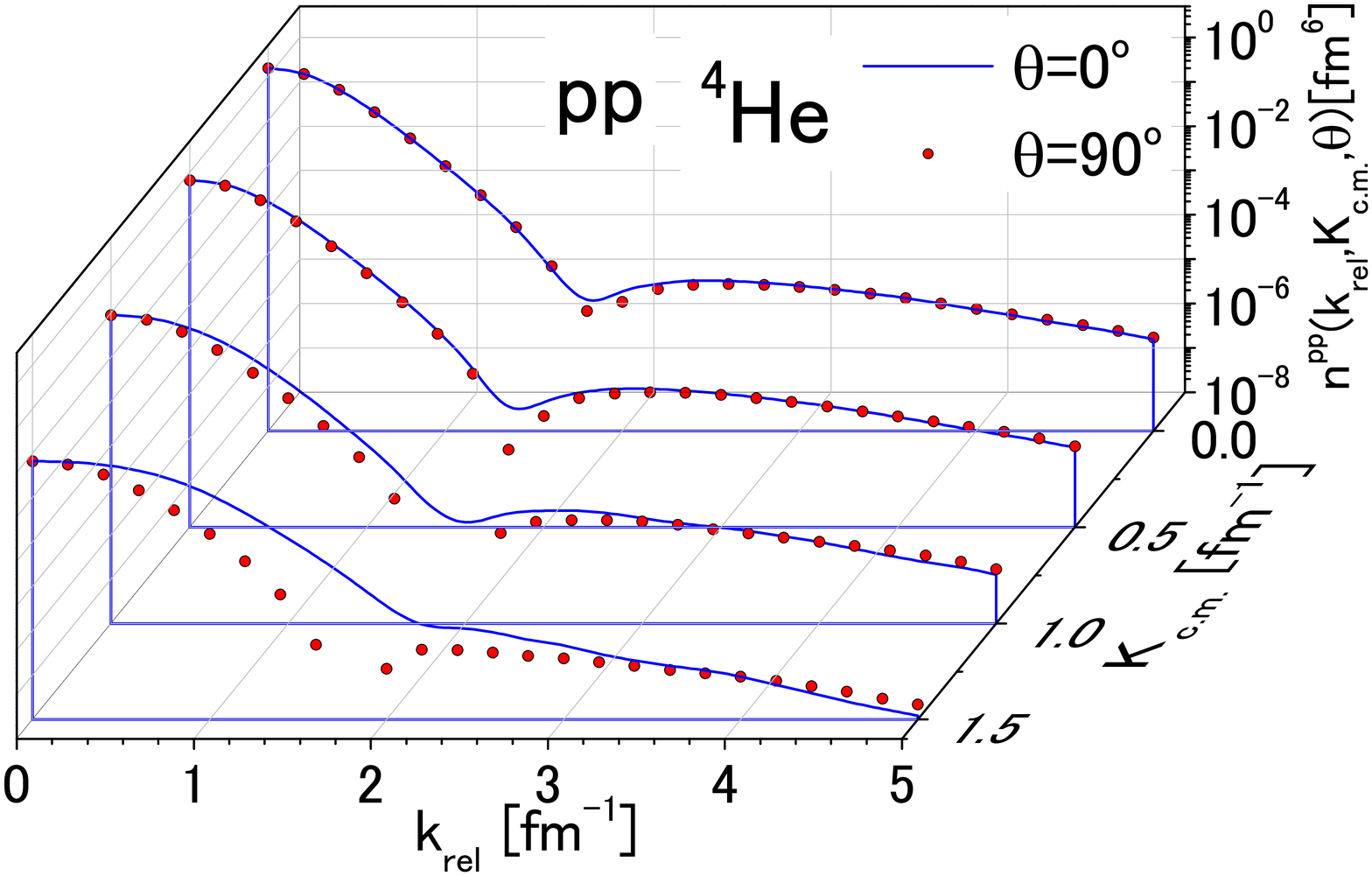} }
 \caption{ (Color online) The $pn$   and $pp$ two-nucleon
momentum distributions  in $^4$He, $n^{pn}(k_{rel}, K_{c.m.},
\theta)$, obtained in Ref. \cite{2:Alvioli:2016} in
correspondence of several values of $K_{c.m.}$ and  two values of
the angle $\theta$ between ${\bf K}_{c.m.}$ and  ${\bf k}_{rel}$.
The region of $k_{rel}$ where the value of $n^{pn}(k_{rel},
K_{c.m.}, \theta)$ is independent of the angle determines the
region of factorization of the momentum distributions, i.e.
$n^{pn}(k_{rel}, K_{c.m.}, \theta) \rightarrow
n_{rel}^{pn}(k_{rel})n^{pn}_{c.m.}(K_{c.m.})$. It can be seen that
the region of factorization starts at  values of $k_{rel}=
k_{rel}^{-}$, which increases with increasing values of
$K_{c.m.}$, i.e.  $k_{rel}^{-} = k_{rel}^{-}(K_{c.m.})$; because
of the dependence of $k_{rel}^{-}$ upon $K_{c.m.}$, a constraint
on the region of
 integration over $K_{c.m.}$  arises from   Eq. (\ref{restriction}).}
 \label{Fig7}
%\end{center}
\end{figure}
\clearpage
%%%%%%%%%%%%%%%%%%%%%%%%%%%%%%%%%%%%%%%%%%%%%%%%%%%%%%%%%%%%%%%%%%%%%%%%%% N E W F I G U R E
\begin{figure}
\begin{center}
\includegraphics[width=0.75\textwidth,keepaspectratio]{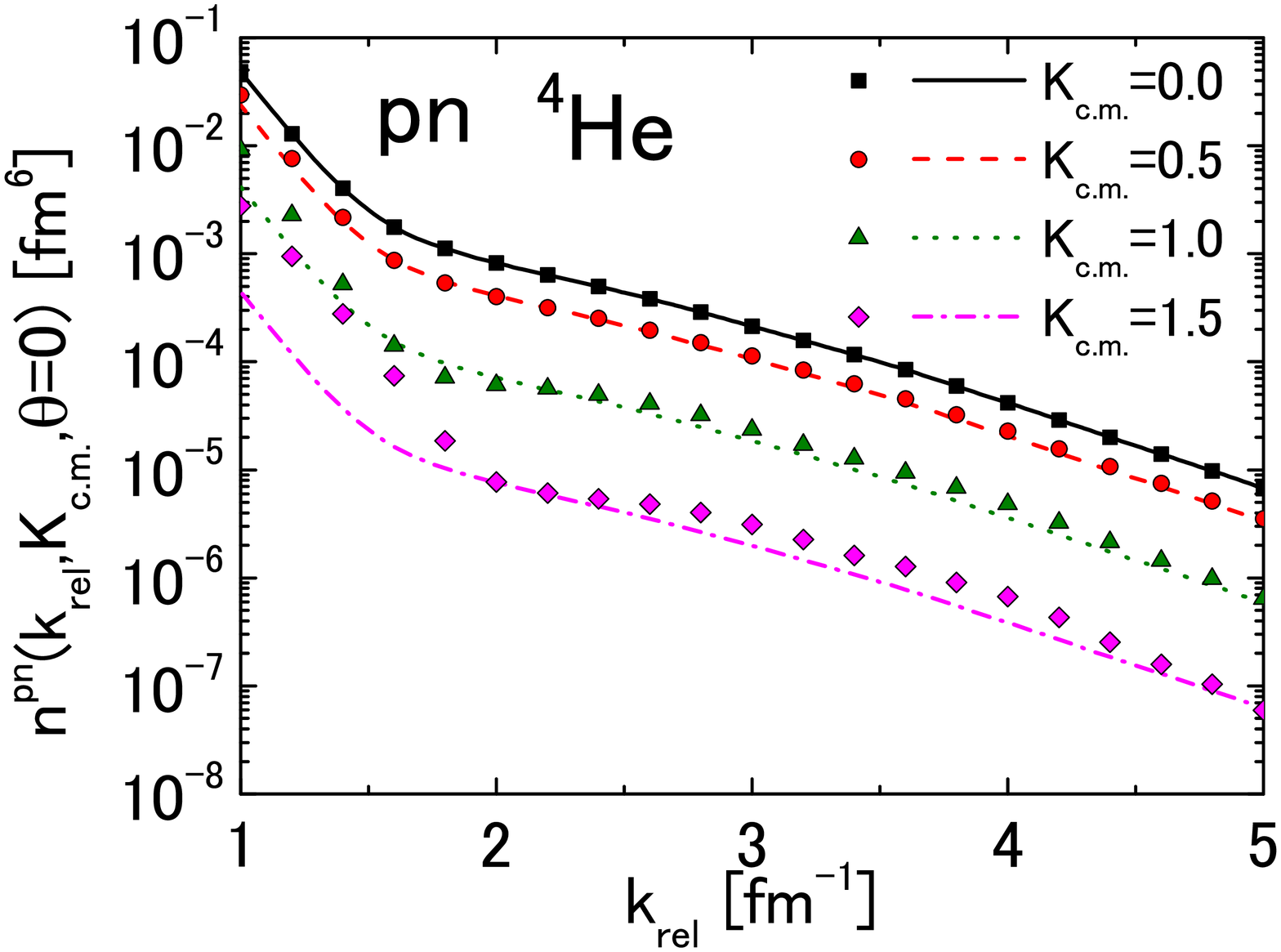}
\vskip -1cm
\includegraphics[width=0.75\textwidth,keepaspectratio]{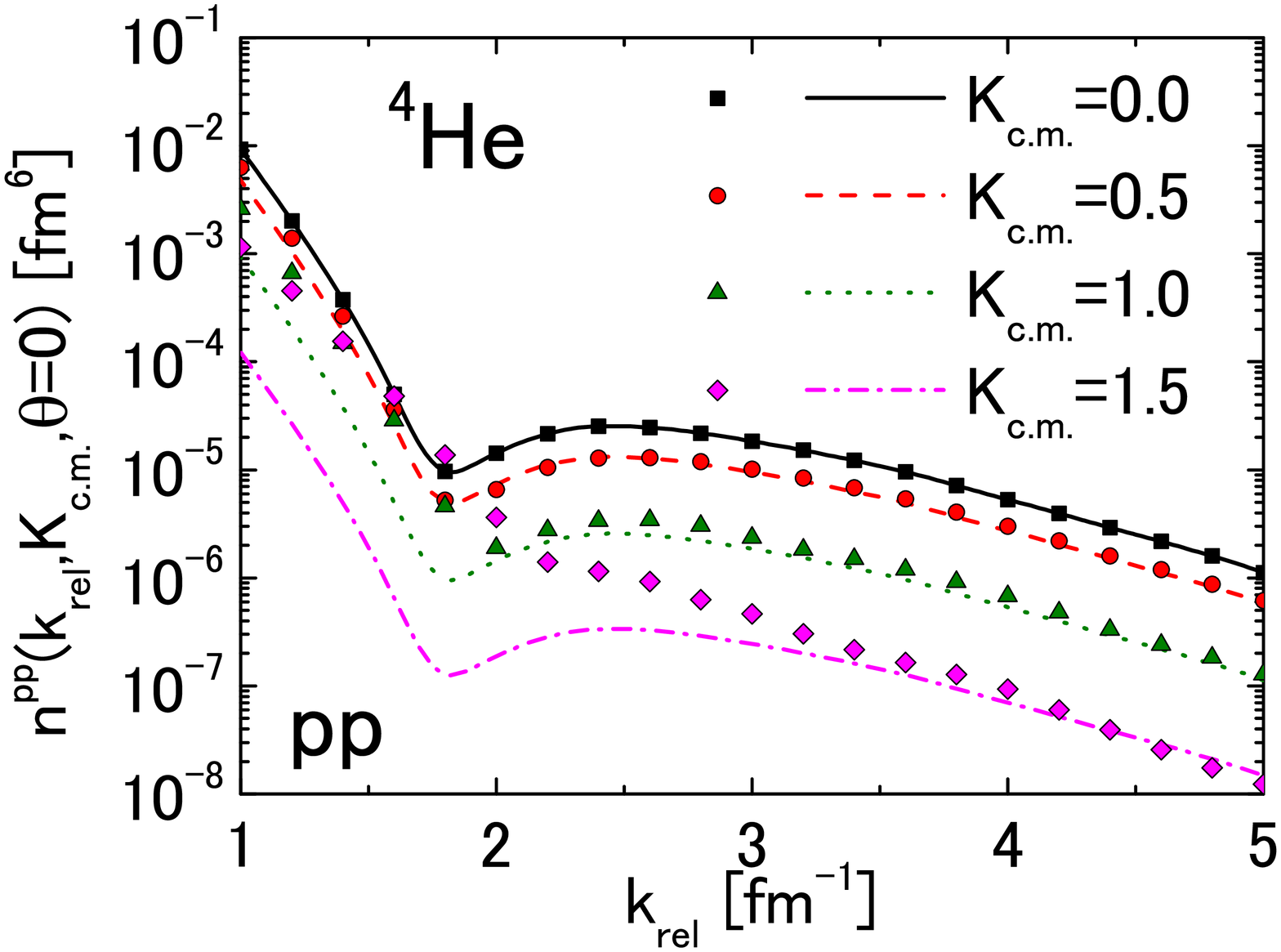}
\end{center}
 %\vskip -1.0cm
 \caption{(Color online)
 The $pn$ and $pp$ two-nucleon momentum distributions
 $n^{pN}(k_{rel}, K_{c.m.}, \theta=0)$ for $^4$He (symbols) compared with the
 results of Eq. (\ref{factorization})  (lines), where for $pn$ and $pp$ Eq. (\ref{factpn}) and Eq.(\ref{ennepp}), respectively have been used.}
\label{Fig8}
%\end{center}
\end{figure}
%%%%%%%%%%%%%%%%%%%%%%%%%%%%%%%%%%%%%%%%%%%%%%%%%%%%%%%%%%%%%%%%%%%%%%%%%% N E W F I G U R E
\newpage
\begin{figure}[ht]
\begin{center}
\includegraphics[width=0.85\textwidth,keepaspectratio]{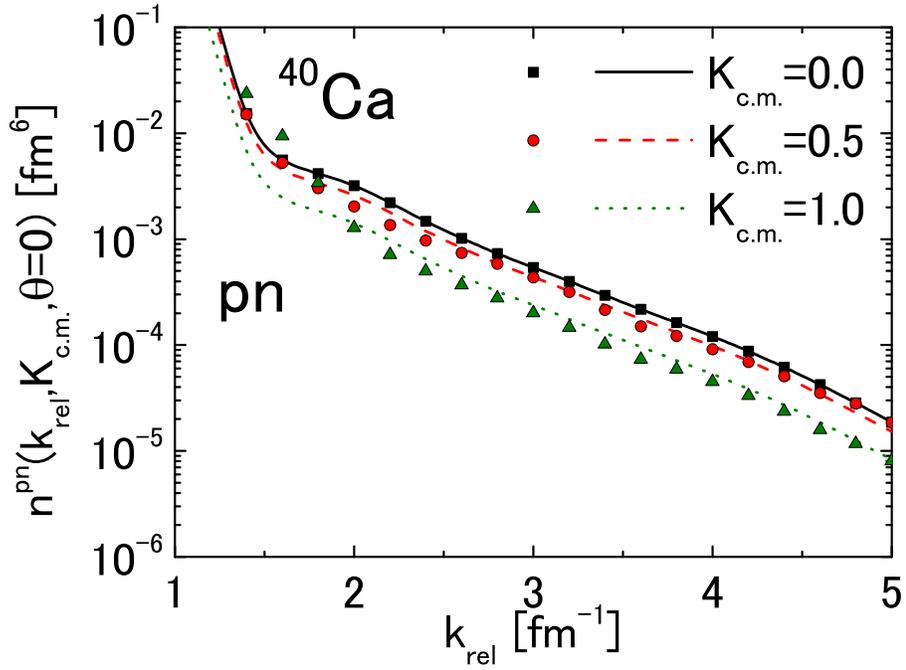}
\includegraphics[width=0.85\textwidth,keepaspectratio]{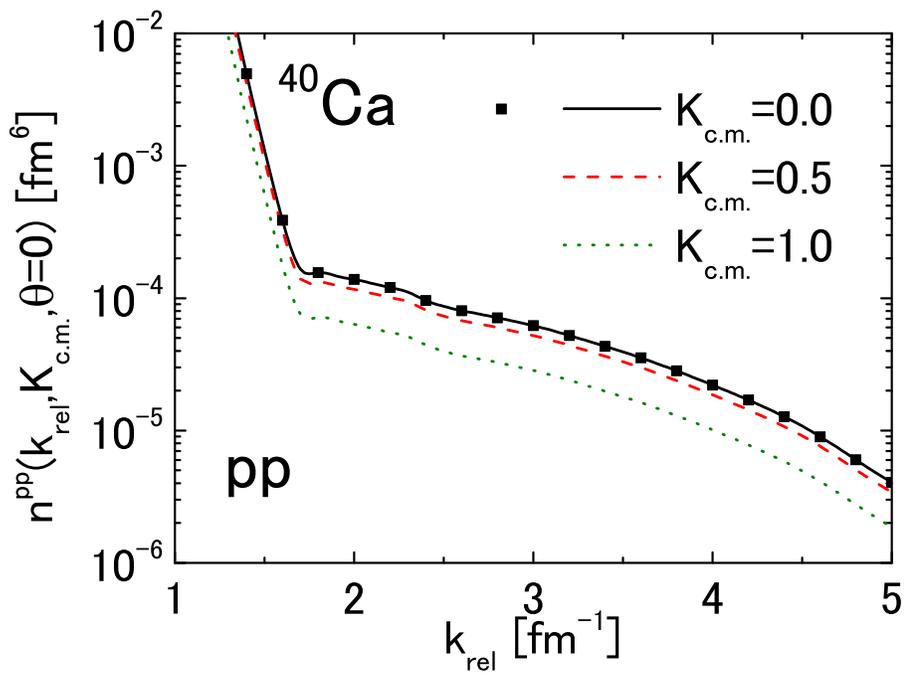}
\vskip 1.0cm
 \caption{(Color online)
 The same as in Fig. \ref{Fig8} but for $^{40}$Ca.}
 \end{center}
\label{Fig9}
\end{figure}
\newpage
\begin{figure}
\centerline{
\includegraphics[width=0.65\textwidth,keepaspectratio]
{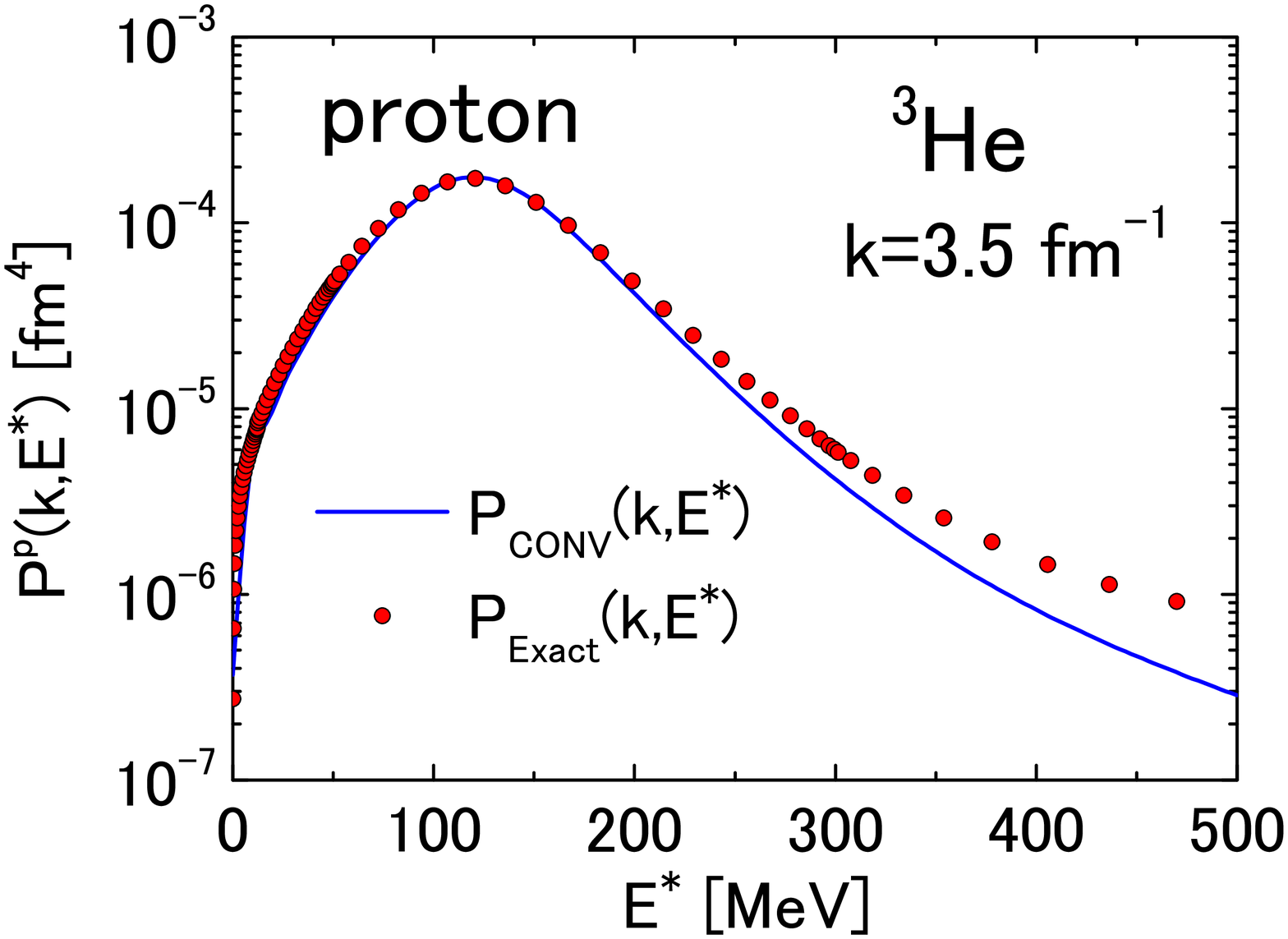}}
\centerline{
\includegraphics[width=0.65\textwidth,keepaspectratio]
{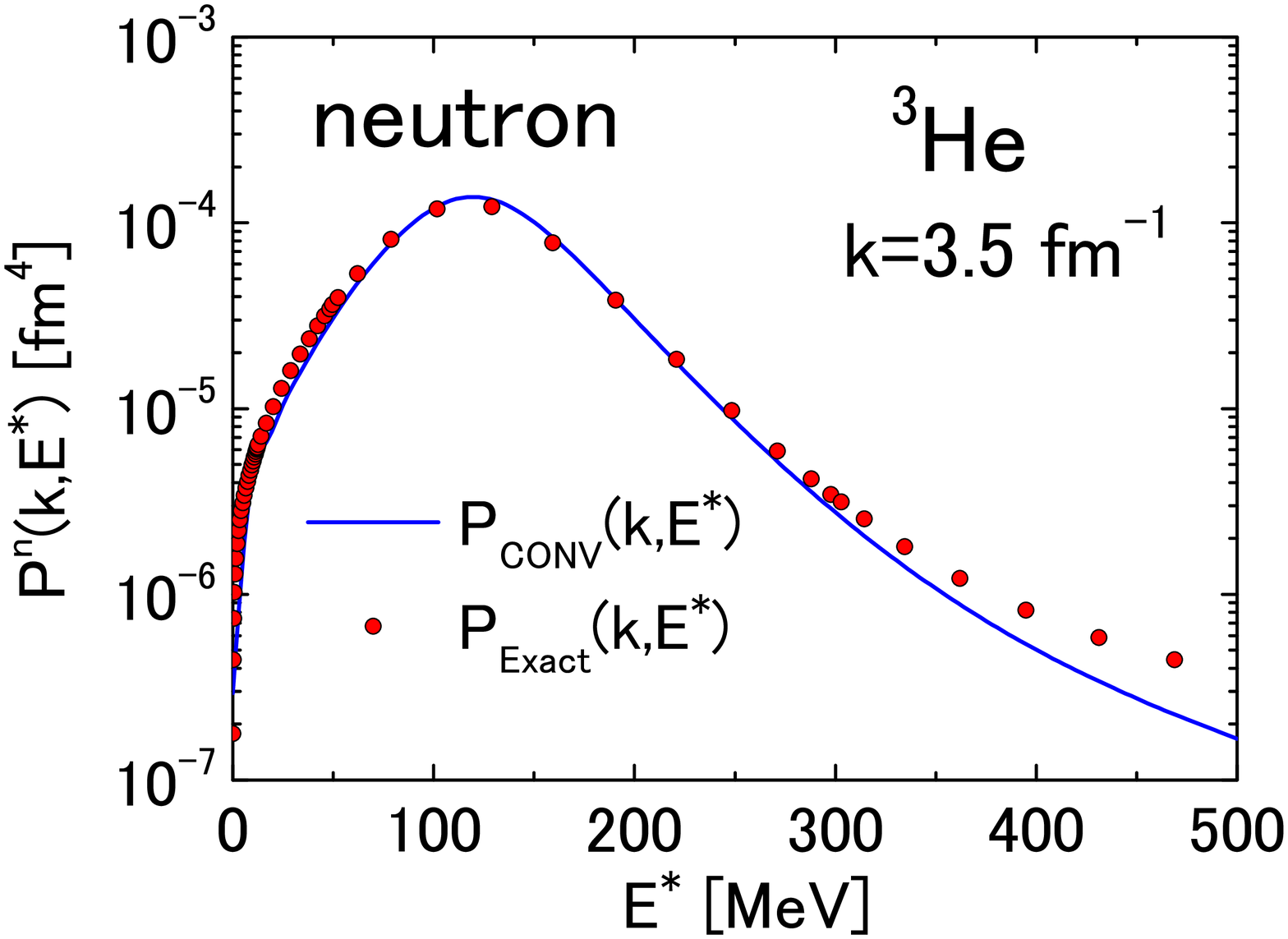}} \caption{(Color online) The {\it
ab-initio} proton and neutron SF of $^3$He from Ref. \cite{Ciofi_Kaptari_Spectral}
 (red dots) compared with  the convolution
SF (Eq. (\ref{Spec_SRC_fin}), full line) obtained
taking into account the constraint (Eq.  (\ref{restriction})) on
the value of $k_{rel}^-$ which guaranties that the convolution
formula includes indeed only the factorization region}
\label{Fig10}
\end{figure}
%%%%%%%%%%%%%%%%%%%%%%%%%%%%%%%%%%%%%%%%%%%%%%%%%%%%%%%%%%%%%%%%%%%%%%%%%% N E W F I G U R E
\begin{figure}
\centerline{
\includegraphics[width=0.6\textwidth,keepaspectratio] {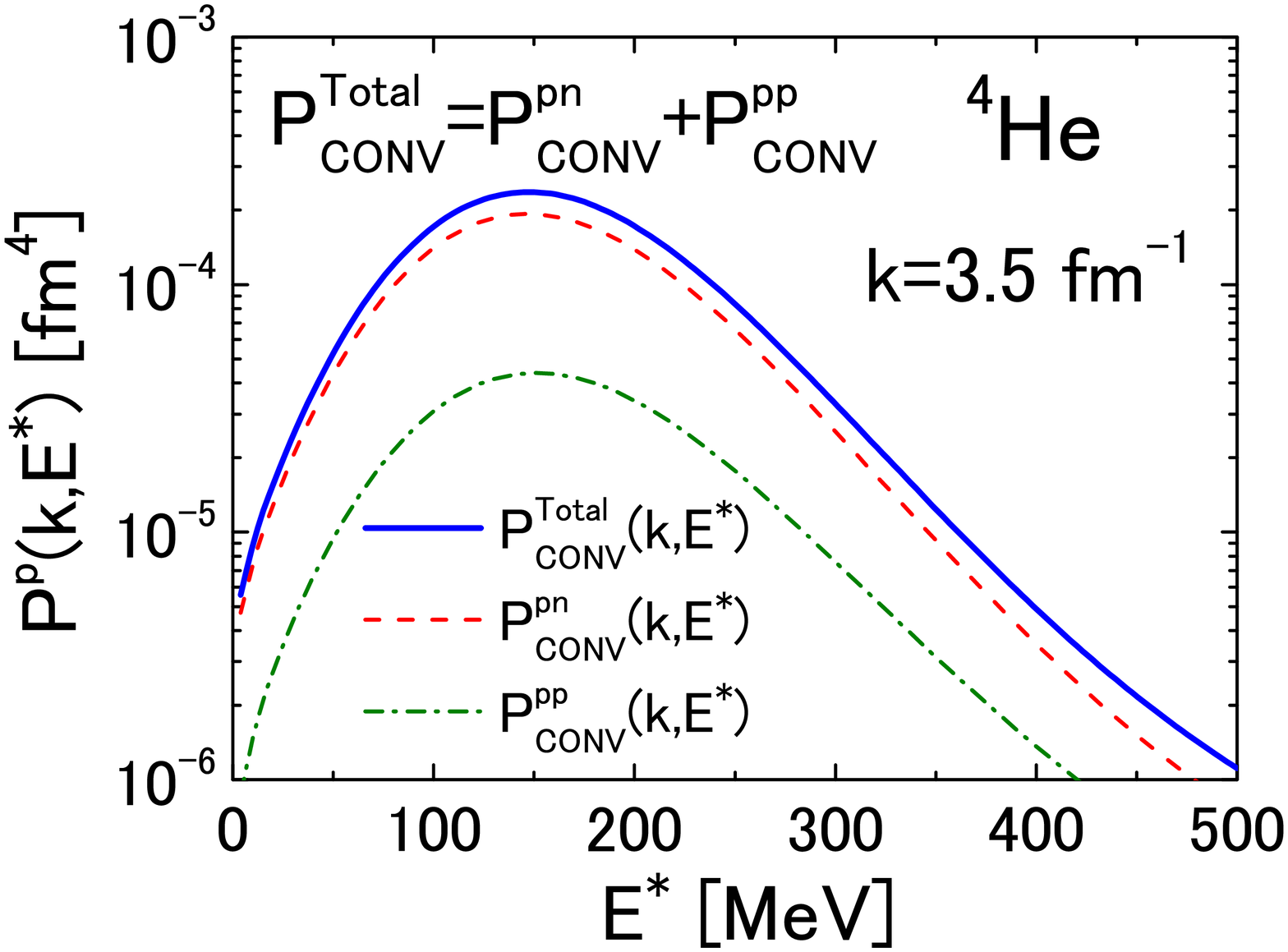}
\includegraphics[width=0.6\textwidth,keepaspectratio]
{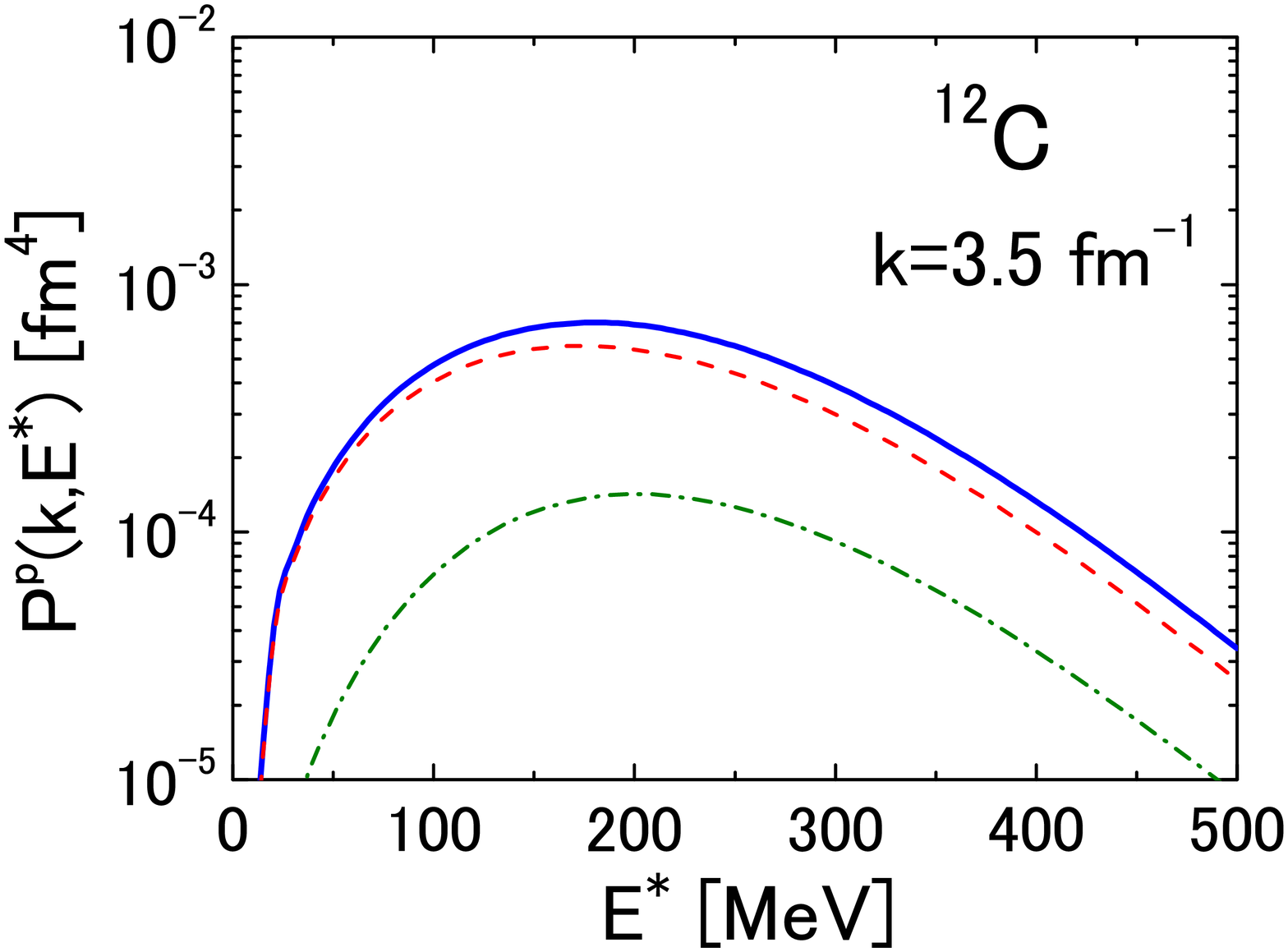}}
%\vskip 1.0cm %\hskip -3cm
\centerline{\includegraphics[width=0.6\textwidth,keepaspectratio]
{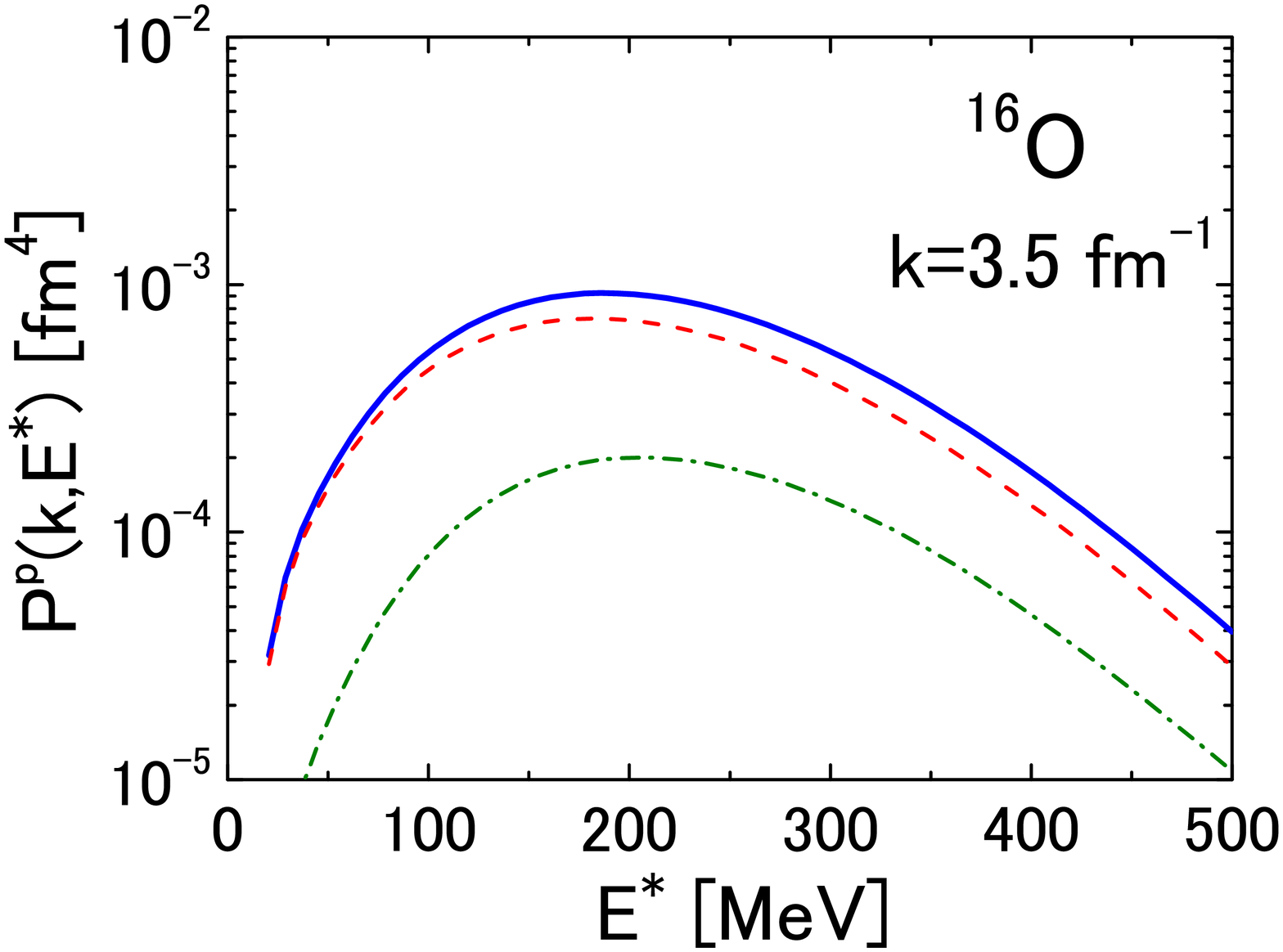}
\includegraphics[width=0.6\textwidth,keepaspectratio]
{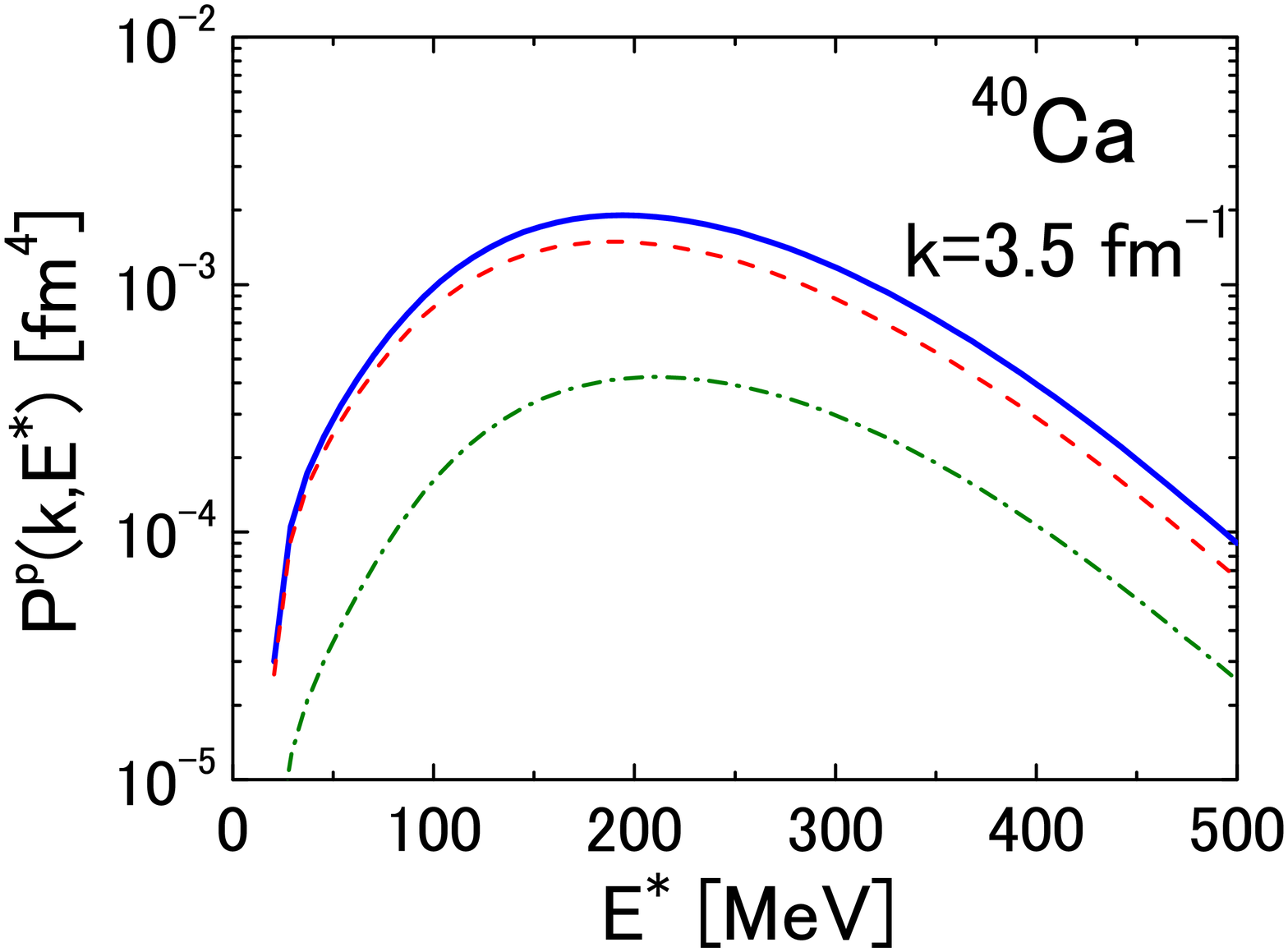}}
%\vskip -0.3cm
\caption{(Color online) The SF of $^4$He, $^{12}$C,
$^{16}$O, and $^{40}$Ca calculated with
 the convolution
SF (Eq. (\ref{Spec_SRC_fin})). The dashed and dot-dashed
lines represent, respectively,
 the $pn$ and
the $pp$ SRC contributions.} \label{Fig11}
\end{figure}
\clearpage
%%%%%%%%%%%%%%%%%%%%%%%%%%%%%%%%%%%%%%%%%%%%%%%%%%%%%%%%%%%%%%%%%%%%%%%%%% N E W F I G U R E
\begin{figure}
\centerline{
\includegraphics[width=0.6\textwidth,keepaspectratio] {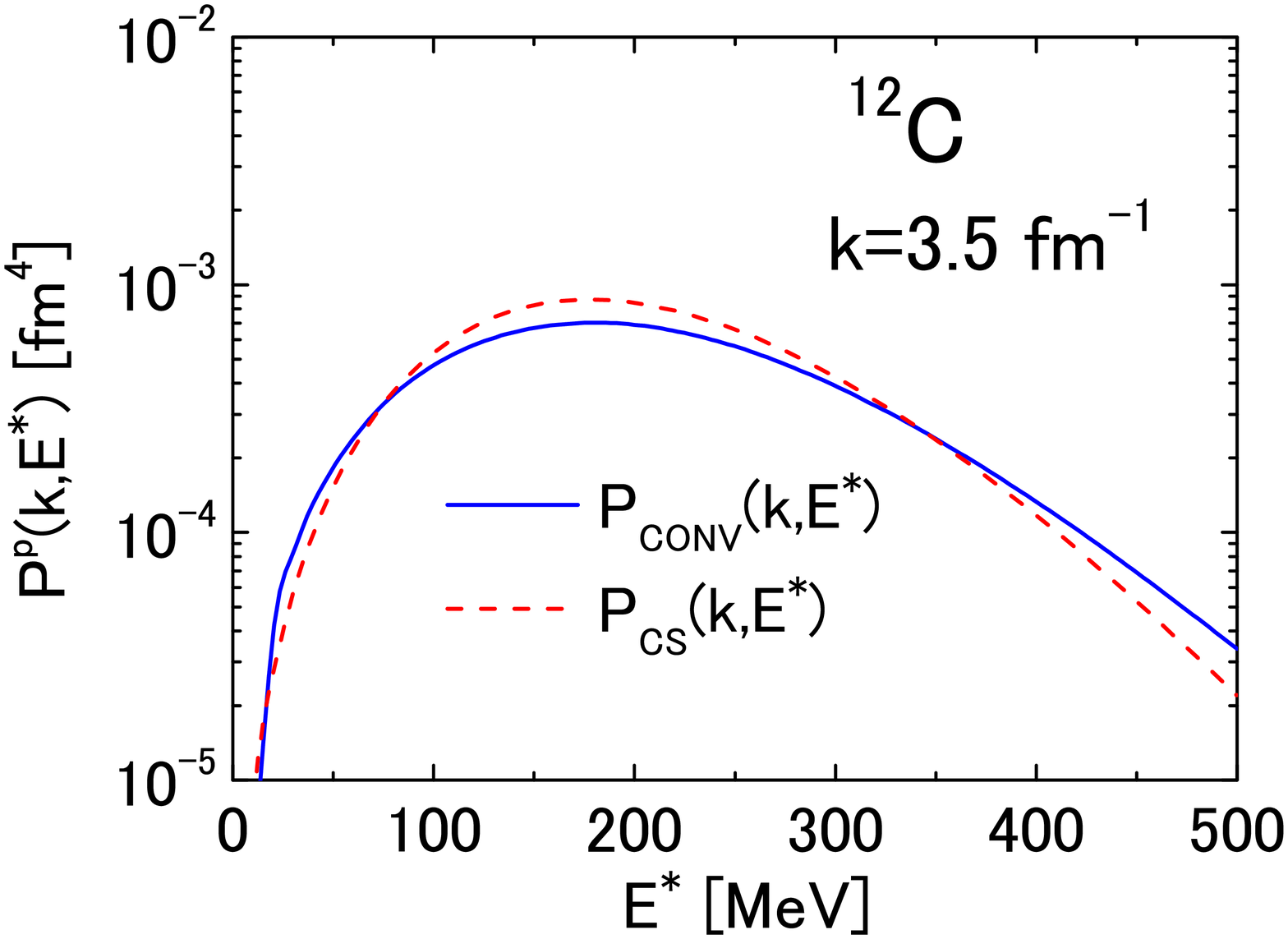}}
\centerline{
\includegraphics[width=0.6\textwidth,keepaspectratio] {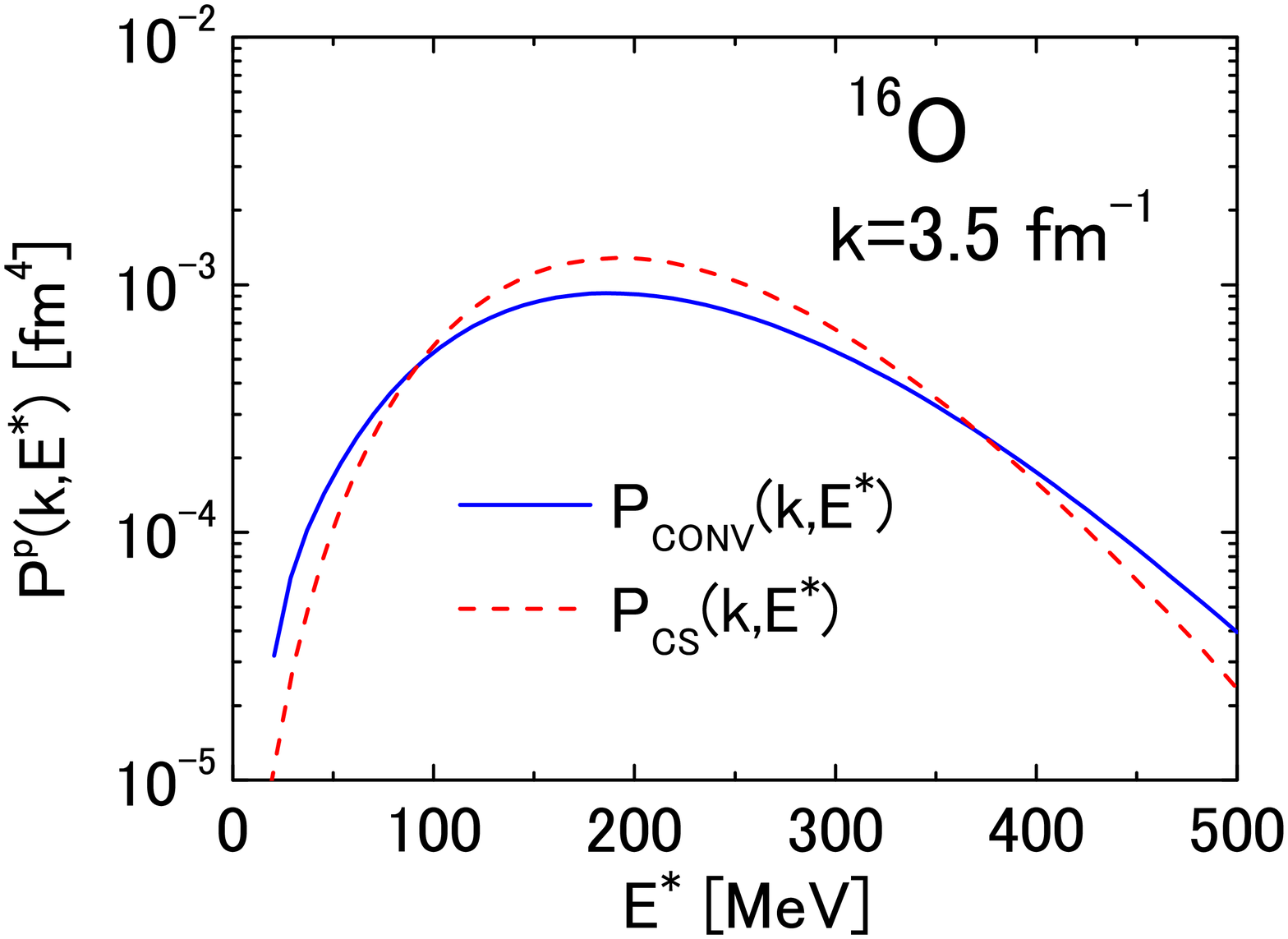}}
\caption{(Color online) The  convolution spectral of $^{12}$C (Eq. (\ref{Spec_SRC_fin})) and
$^{16}$O (full line)
  compared with the effective convolution formula from
Ref. \cite{CiofidegliAtti:1995qe} (dashed line).}
 \label{Fig12}
\end{figure}
%%%%%%%%%%%%%%%%%%%%%%%%%%%%%%%%%%%%%%%%%%%%%%%%%%%%%%%%%%%%%%%%%%%%%%%%%% N E W F I G U R E
\clearpage
\begin{figure}
\centerline{
\includegraphics[width=0.7\textwidth,keepaspectratio] {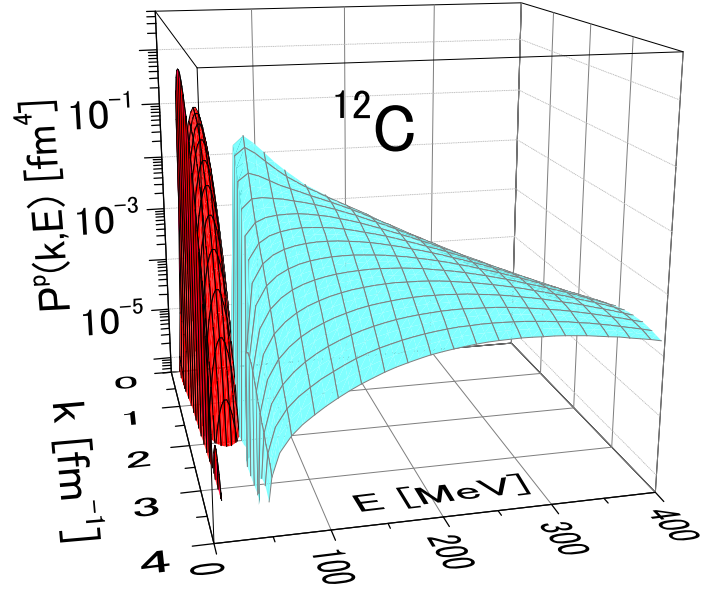}}
\caption{(Color online) A 3D figure of the total SF (Eq. (\ref{Spec_fin})) of
 $^{12}C$ illustrating the mean-field and SRC contributions: $P_{MF}^{N_1}(k,E)$ (shown in Red) is located in
 the region of removal energy $E \leq 50 \, MeV$  where  the contribution from the $\ell=1$ and $\ell=0$
 shells can be identified, whereas $P_{SRC}^{N_1}(k,E)$ (shown in Blue) completely exhausts the removal energy region with $E \geq 50 \, MeV$.}
\label{Fig13}
\end{figure}

\clearpage
\begin{figure}
\centerline{
\includegraphics[width=0.65\textwidth,keepaspectratio] {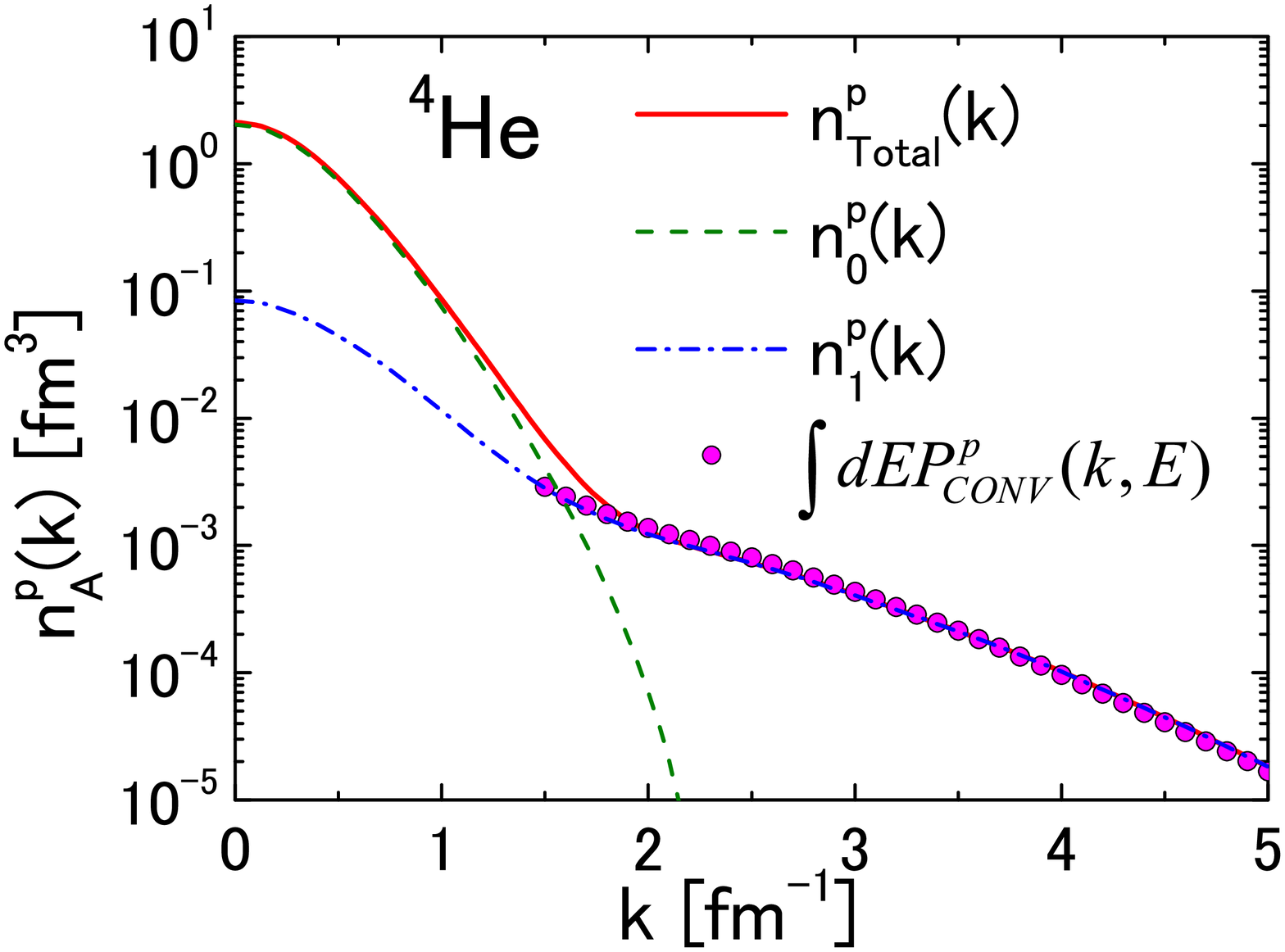}}
\centerline{
\includegraphics[width=0.65\textwidth,keepaspectratio] {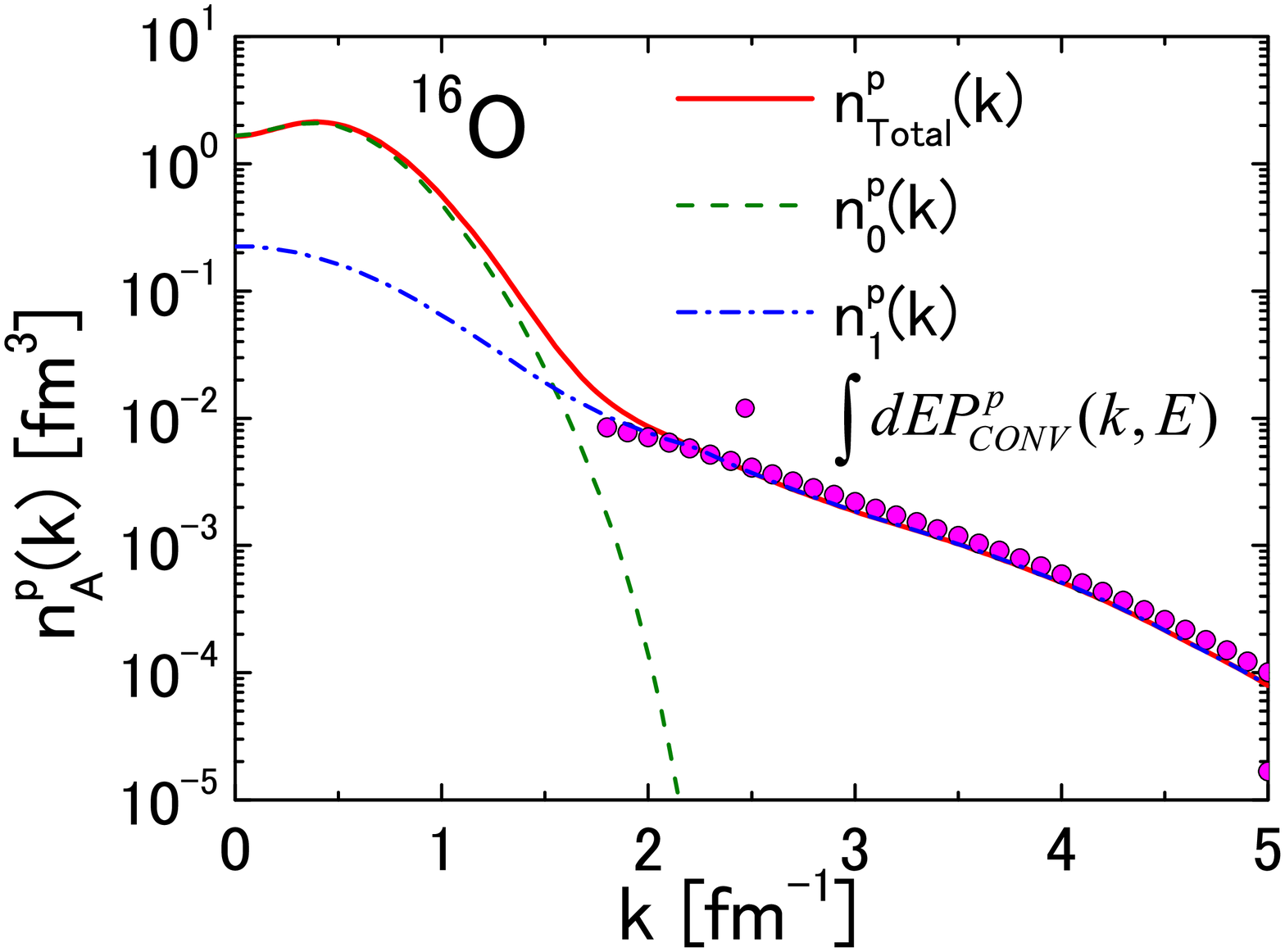}}
\caption{(Color online) {\it The SRC Momentum sum rule
$n_{SRC}(k) \equiv n_1(k) = \int_0^{\infty} P(k,E^*)dE^*$ in $^4$He and $^{16}$O}.
 The full line represents the total momentum distribution
obtained in Ref. \cite{2:Alvioli:2005cz}  with the dashed and dot-dashed curves corresponding
to the mean-field and SRC contributions, respectively. The full dots represent the the  SRC momentum
 distribution obtained by integrating the
 the SRC  convolution SF. It can be seen that the momentum sum rule is
 exactly satisfied by the convolution formula.}
 \label{Fig14}
\end{figure}
%%%%%%%%%%%%%%%%%%%%%%%%%%%%%%%%%%%%%%%%%%%%%%%%%%%%%%%%%%%%%%%

%%%%%%%%%%%%%%%%%%%%%%%%%%%%%%%%%%%%%%%%%%%%%%%%%%%%%%%%%%%%%%%%%%%%%%%%%% N E W F I G U R E
\clearpage
\begin{figure}
\centerline{
\includegraphics[width=0.65\textwidth,keepaspectratio] {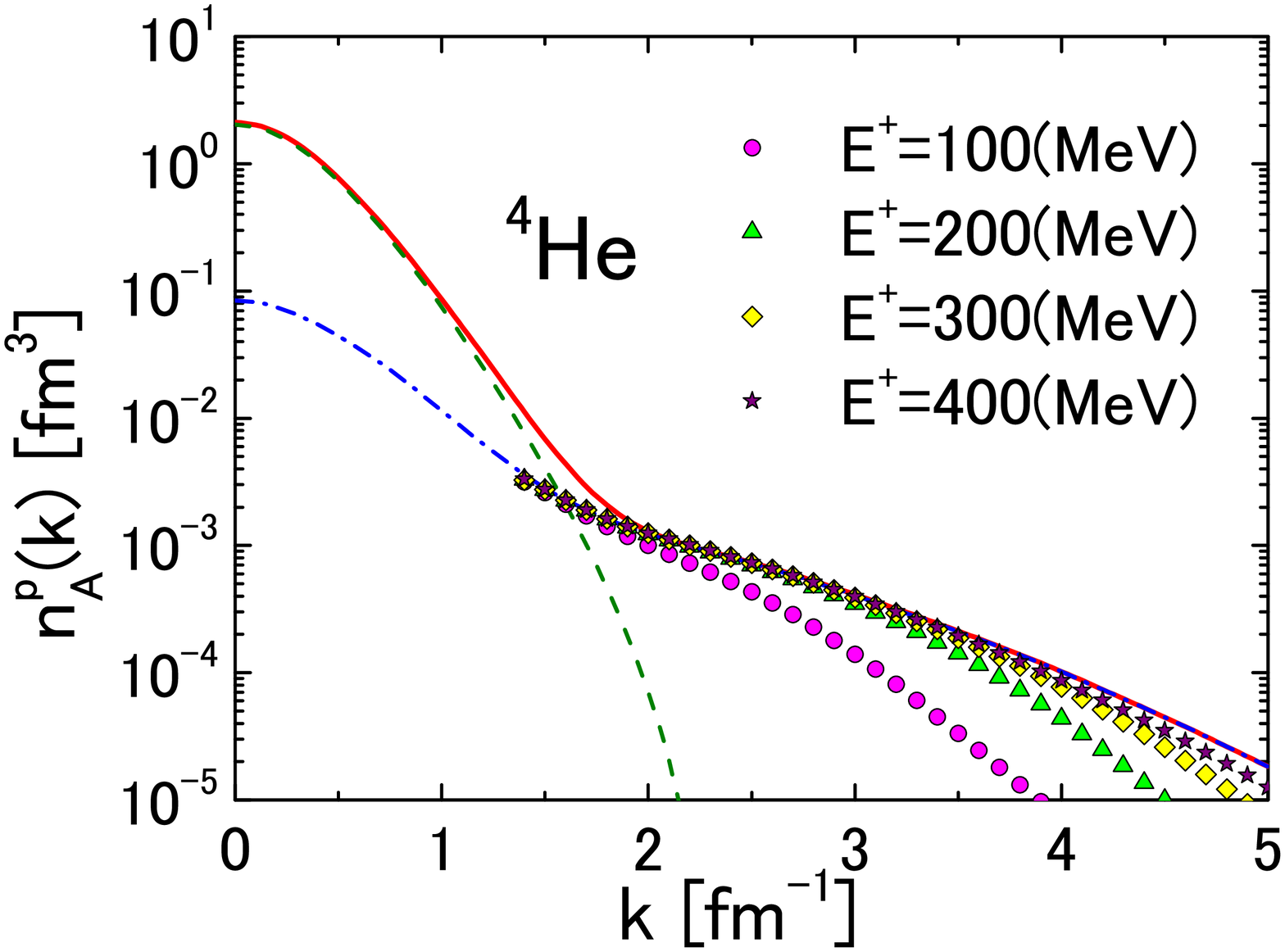}}
\centerline{
\includegraphics[width=0.65\textwidth,keepaspectratio] {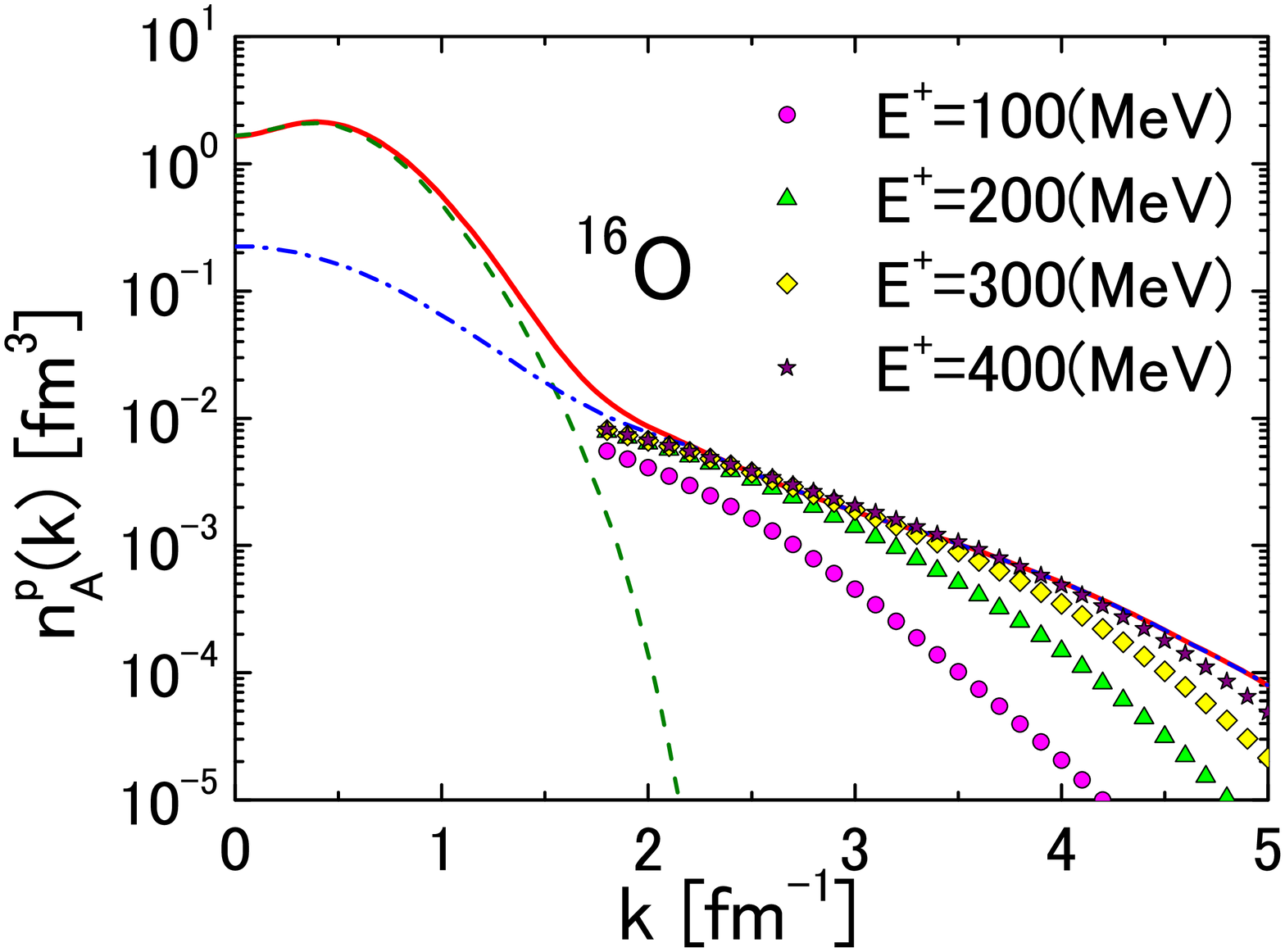}}
\caption{(Color online) {\it  The convergence of the momentum sum
rule $n_{SRC}(k)\equiv n_1(k)=\int_0^{E^+} P(k,E^*)\,d\,E^*$.} The
partial momentum sum rule corresponding to  increasing value
of $E^{+}$. It can be seen that in order to obtain the correct
momentum distributions in the region $k\geq 4 \, fm^{-1}$ it is
necessary to integrate the SF up to $E^{+} \simeq
400 \, MeV$. Full, dashed and dot-dashed curves as in Fig. \ref{Fig14}.} \label{Fig15}
\end{figure}
\end{document}